\documentclass[12pt,draftcls,journal,onecolumn]{IEEEtran}
\usepackage{graphicx}
\usepackage{epstopdf}
\usepackage[latin1]{inputenc}
\usepackage{amsmath,amssymb,amscd,latexsym,dsfont}
\usepackage[english]{babel}
\usepackage{tabularx,cite}
\usepackage{graphicx}
\usepackage{algpseudocode}
\usepackage{algorithm}
\usepackage{algorithmicx}
\usepackage{float}
\usepackage{multicol}
\usepackage{psfrag}
\usepackage{comment,psfrag,subfigure,enumerate}
\usepackage{color}
\usepackage{amsthm}
\usepackage{graphicx}
\usepackage{graphicx}
\usepackage{epstopdf}
\usepackage[latin1]{inputenc}
\usepackage{amsmath,amssymb,amscd,latexsym,dsfont}
\usepackage[english]{babel}
\usepackage{tabularx,cite}
\usepackage{graphicx}
\usepackage{algpseudocode}
\usepackage{algorithm}
\usepackage{algorithmicx}
\usepackage{float}
\usepackage{multicol}
\usepackage{mathtools}
\usepackage{psfrag}
\usepackage{comment,psfrag,subfigure,enumerate}
\usepackage{color}
\usepackage{amsthm}
\ifCLASSINFOpdf
\else
\fi
\begin{document}
\IEEEoverridecommandlockouts
\IEEEpubid{\makebox[\columnwidth]{978-1-4799-5863-4/14/\$31.00 \copyright 2014 IEEE \hfill} \hspace{\columnsep}\makebox[\columnwidth]{ }}
\title{On the Performance of Millimeter Wave-based RF-FSO Multi-hop and Mesh Networks}

\author{
    \IEEEauthorblockN{Behrooz Makki\IEEEauthorrefmark{1}, Tommy Svensson\IEEEauthorrefmark{1}, \emph{Senior Member, IEEE,} Maite Brandt-Pearce\IEEEauthorrefmark{2} \emph{Senior Member, IEEE,} and Mohamed-Slim Alouini\IEEEauthorrefmark{3}, \emph{Fellow, IEEE}}

    \IEEEauthorblockA{\IEEEauthorrefmark{1}Chalmers University of Technology, Gothenburg, Sweden, \{behrooz.makki, tommy.svensson\}@chalmers.se}

\IEEEauthorblockA{\IEEEauthorrefmark{2}University of Virginia, Charlottesville, VA , USA, mb-p@virginia.edu}

    \IEEEauthorblockA{\IEEEauthorrefmark{3} King Abdullah University of Science and Technology (KAUST), Thuwal, Saudi Arabia, slim.alouini@kaust.edu.sa}
    \thanks{Part of this work has been accepted for presentation at the IEEE WCNC 2017.}
}

%
\maketitle
\vspace{-5mm}
\begin{abstract}
This paper studies the performance of multi-hop and mesh networks composed of millimeter wave (MMW)-based radio frequency (RF) and free-space optical (FSO) links. The results are obtained in cases with and without hybrid automatic repeat request (HARQ). Taking the MMW characteristics of the RF links into account, we derive closed-form expressions for the networks' outage probability and ergodic achievable rates. We also evaluate the effect of various parameters such as power amplifiers efficiency, number of antennas as well as different coherence times of the RF and the FSO links on the system performance. Finally, we determine the minimum number of the transmit antennas in the RF link such that the same rate is supported in the RF- and the FSO-based hops. The results show the efficiency of the RF-FSO setups in different conditions. Moreover, HARQ can effectively improve the outage probability/energy efficiency, and compensate for the effect of hardware impairments in RF-FSO networks. For common parameter settings of the RF-FSO dual-hop networks, outage probability of $10^{-4}$ and code rate of $3$ nats-per-channel-use, the implementation of HARQ with a maximum of $2$ and $3$ retransmissions reduces the required power, compared to cases with open-loop communication, by $13$ and $17$ dB, respectively.
\end{abstract}



%
\IEEEpeerreviewmaketitle
\vspace{-0mm}
\section{Introduction}
The next generation of wireless networks must provide coverage for everyone everywhere at any time. To address these demands, a combination of different techniques {is} considered, among which free-space optical (FSO) communication is very promising  \cite{6331134,6932439,6887284}. Coherent FSO systems, made inexpensive by the large fiberoptic market, provide fiber-like data rates through the atmosphere using lasers. Thus, FSO can be used for a wide range of applications such as last-mile access, fiber back-up, back-hauling and multi-hop networks. In the radio frequency (RF) domain, on the other hand,  millimeter wave (MMW) communication has emerged as a key enabler to obtain sufficiently large bandwidths so that it is possible to achieve data rates comparable to those in the FSO links. In this perspective, the combination of FSO and MMW-based RF links is considered as a powerful candidate for high-rate reliable communication.

The RF-FSO related {literature} can be divided into two {groups}. The first group {consists of} papers on single-hop setups where the link reliability is improved via the joint implementation of RF and FSO systems. Here, either the RF and the FSO links are considered as separate links and the RF link acts as a backup when the FSO link is down, e.g., \cite{6887284,4610745,1399401,4168193,4393998,Hamzeh,6364576,6400459}, or the links are combined to improve the system performance \cite{6503564,5342330,4411336,4348339,6222288,5351671,5427418}.  Also, the implementation of hybrid automatic repeat request (HARQ) in RF-FSO links has been considered in \cite{5427418,6692504,7445896}.

The second group {consists of} the papers analyzing the performance of multi-hop RF-FSO systems. For instance, \cite{6831655,6866170} study RF-FSO based relaying schemes with an RF source-relay link and an FSO or RF-FSO relay-destination link. Also, considering Rayleigh fading conditions for the RF link and amplify-and-forward relaying technique, \cite{6678140,5999707} derive the end-to-end error probability of the RF-FSO based setups and compare the system performance with RF-based relay networks, respectively. The impact of pointing errors on the performance of dual-hop RF-FSO systems is studied in \cite{7127443,6512100,7055847}. Finally, \cite{6775014} analyzes decode-and-forward techniques in multiuser relay networks using RF-FSO.

In this paper, we study the data transmission efficiency of multi-hop and mesh RF-FSO systems from an information theoretic point of view.
Considering the MMW characteristics of the RF links and heterodyne detection technique in the FSO links, we derive closed-form expressions for the system outage probability  (Lemmas 1-6) and ergodic achievable rates (Corollary 2). Our results are obtained for the decode-and-forward relaying approach in different cases with and without HARQ. Specifically, we show the HARQ as an effective technique to compensate for the non-ideal properties of the RF-FSO system and improve the network reliability. We present mappings between the performance of RF- and FSO-based hops  as well as between the HARQ-based and open-loop systems, in the sense that with appropriate parameter settings the same outage probability is achieved in these setups (Corollary 1, Lemma 6). Also, we determine the minimum number of transmit antennas in the RF links such that the same rate is supported by the RF- and the FSO-based hops (Corollary 2). Finally, we analyze the effect of various parameters such as the power amplifiers (PAs) efficiency, different coherence times of the RF and FSO links and number of transmit antennas on the performance of multi-hop and mesh networks.

In contrast to \cite{6331134,6932439,6887284,4610745,1399401,4168193,4393998,Hamzeh,6364576,6400459,6503564,6222288,5342330,4411336,4348339,5351671,5427418,6692504,7445896}, we consider multi-hop and mesh networks. Moreover,  
our  analytical/numerical results on the outage probability, ergodic achievable rate  and the required number of antennas in HARQ-based RF-FSO systems as well as our discussions on the effect of imperfect PAs/HARQ have not been presented before. The differences in the problem formulation and the channel model makes our analytical/numerical results and conclusions completely different from the ones in the literature, e.g., \cite{6331134,6932439,6887284,4610745,1399401,4168193,4393998,Hamzeh,6364576,6400459,6503564,5342330,4411336,4348339,5351671,5427418,6692504,7445896,6831655,6866170,6678140,5999707,7127443,6512100,7055847,6775014,6222288}.

The numerical and the analytical results show that:
\begin{itemize}
  \item Depending on the codewords length, there are different methods for the analytical performance evaluation of the RF-FSO systems (Lemmas 1-6).
  \item There are mappings between the performance of RF- and FSO-based hops, in the sense that with proper scaling of the channel parameters the same outage probability is achieved in these hops (Corollary 1). Thus, the performance of RF-FSO based multi-hop/mesh networks can be mapped to ones using only the RF- or the FSO-based communication.
  \item While the network outage probability is (almost) insensitive to the number of RF-based transmit antennas  when this number is large, the ergodic rate of the multi-hop network is remarkably affected by the number of antennas.
  \item The required number of RF-based antennas to guarantee the same rate as in the FSO-based hops increases significantly with the signal-to-noise ratio (SNR) and, at high SNRs, the ergodic rate scales with the SNR (almost) linearly.
  \item At low SNRs, the same outage probability is achieved in  HARQ-based RF hops with $N$ transmit antennas, a maximum of $M$ retransmissions and $C$ channel realizations per retransmission as with  an open-loop system with $MNC$ transmit antennas and single channel realization per codeword transmission (Lemma 6).
  \item The PAs efficiency affects the network outage probability/ergodic rate considerably. However, the HARQ protocols can effectively  compensate for the effect of hardware impairments.
  \item Finally, the HARQ improves the outage probability/energy efficiency significantly. For instance, consider common parameter settings of the RF-FSO dual-hop networks, outage probability of $10^{-4}$ and code rate of $3$ nats-per-channel-use (npcu). Then, compared to cases with open-loop communication, the implementation of HARQ with a maximum of $2$ and $3$ retransmissions reduces the required power by $13$ and $17$ dB, respectively.
\end{itemize}
\vspace{-0mm}
\section{System Model}
In this section, we present the system model for a multi-hop setup with a single route from the source to the destination. As demonstrated in Section III.C, the results of the multi-hop networks can be extended to the ones in mesh networks with multiple non-overlapping routes from the source to the destination.
\subsection{Channel Model}
Consider a $T^\text{total}$-hop RF-FSO system, with $T$ RF-based hops and $\tilde T=T^\text{total}-T$ FSO-based hops. As seen in the following, the outage probability and the ergodic achievable rate are independent of the order of the hops. Thus, we do not need to specify the order of the RF- and FSO-based hops. The $i$-th, $i=1,\ldots,T,$ RF-based hop uses a multiple-input-single-output (MISO) setup with  $N_{i}$ transmit antennas. Such a setup is of interest in, e.g., side-to-side communication between buildings/lamp posts \cite{mmmagicdeliverable}, as well as in wireless backhaul links where the trend is to introduce multiple antennas and thereby {achieve} multiple parallel streams, e.g., \cite{ericssonAB}. We define the channel gains as $g_{i}^{j_i}\doteq|h_{i}^{j_i}|^2,i=1,\ldots, T,j_i=1,\ldots,N_i,$ where $h_i^{j_i}$ is the complex fading coefficients of the channel between the $j_i$-th antenna in the $i$-th hop and its corresponding receive antenna.

While the modeling of the MMW-based links is well known for line-of-sight wireless backhaul links, it is still an ongoing research topic  for non-line-of-sight conditions \cite{mmmagic}. Particularly, different measurement setups have emphasized the near-line-of-sight propagation and the non-ideal hardware as two key features of such links. Here, we present the analytical results for the quasi-static Rician channel model, with successive independent realizations, which is an appropriate model for near line-of-sight conditions and has been well established for different MMW-based applications, e.g., \cite{mmwrician1,mmwrician2,mmwrician3}.

Let us denote the probability density function (PDF) and the cumulative distribution function (CDF) of a random variable $X$ by $f_X(\cdot)$ and $F_X(\cdot)$, respectively. With a Rician model, the channel gain ${g_i^{j_i},\forall i,j_i,}$ follows the PDF
\begin{align}\label{eq:eqRicianpdf}
&f_{g_i^{j_i}}(x)=\frac{(K_i+1)e^{-K_i}}{\Omega_i}e^{-\frac{(K_i+1)x}{\Omega_i}}I_0\left(2\sqrt{\frac{K_i(K_i+1)x}{\Omega_i}} \right ),\forall i,j_i,
\end{align}
where $K_i$ and $\Omega_i$ denote the fading parameters in the $i$-th hop and $I_n(\cdot)$ is the $n$-th order modified Bessel function of the first kind.
Also, defining the sum channel gain $G_i=\sum_{j_i=1}^{N_i}{g_i^{j_i}},$ we have
\begin{align}\label{eq:eqRiciansumpdf}
f_{G_i}(x)&=\frac{(K_i+1)e^{-K_iN_i}}{\Omega_i}\left(\frac{(K_i+1)x}{K_iN_i\Omega_i}\right)^{\frac{N_i-1}{2}}e^{-\frac{(K_i+1)x}{\Omega_i}}I_{N_i-1}\left(2\sqrt{\frac{K_i(K_i+1)N_ix}{\Omega_i}} \right ),\forall i.
\end{align}

Finally, to take the non-ideal hardware into account, we consider the state-of-the-art model for the PA efficiency where the output power at each antenna of the $i$-th hop is determined according to
\cite[eq. (2.14)]{phdthesisBjornemo}, \cite[eq. (3)]{6515206}, \cite[eq. (3)]{6725577}, \cite[eq. (1)]{7104158}
\begin{align}\label{eq:ampmodeldaniel}
&\frac{P_i}{P_i^{\text{cons}}}=\epsilon_i\left(\frac{P_i}{P_i^{\text{max}}}\right)^\vartheta_i\Rightarrow  P_i=\sqrt[1-\vartheta_i]{\frac{\epsilon_i P_i^{\text{cons}}}{(P_i^{\text{max}})^\vartheta_i}},\forall i.
\end{align}
Here, $ P_i, P_i^{\text{max}}$ and $P_i^{\text{cons}},\forall i,$ are the output, the maximum output and the consumed power in each antenna of the $i$-th hop, respectively, $\epsilon_i\in [0,1]$ denotes the maximum power efficiency achieved at $P_i=P_i^{\text{max}}$ and $\vartheta_i\in [0,1]$ is a parameter depending on the PA class.

The FSO links, on the other hand, are assumed to have single transmit/receive terminals. Reviewing the literature and depending on the channel condition, the FSO link may follow different distributions. Here, we present the results for cases with exponential and Gamma-Gamma distributions of the FSO links. For the exponential distribution of the $i$-th FSO hop, the channel gain $\tilde G_i$ follows
\begin{align}\label{eq:Eqexpfsopdf}
f_{\tilde G_i}(x)=\lambda_i e^{-\lambda_i x}, \forall i,
\end{align}
 with $\lambda_i$ being the long-term channel coefficient of the $i$-th, $i=1,\ldots,\tilde T,$ hop. Moreover, with the Gamma-Gamma distribution we have
\begin{align}\label{eq:eqpdfgammagamma}
f_{\tilde G_i}(x)=\frac{2(a_ib_i)^{\frac{a_i+b_i}{2}}}{\Gamma(a_i)\Gamma(b_i)}x^{\frac{a_i+b_i}{2}-1}\mathcal{K}_{a_i-b_i}\left(2\sqrt{a_ib_ix}\right),\forall i.
\end{align}
Here, $\mathcal{K}_n(\cdot)$ denotes the modified Bessel function of the second kind of order $n$ and $\Gamma(x)=\int_0^\infty{u^{x-1}e^{-u}\text{d}u}$ is the Gamma function. Also, $a_i$ and $b_i, i=1,\ldots,\tilde T,$ are the distribution shaping parameters which can be expressed as functions of the Rytov variance, e.g., \cite{7445896}.
\subsection{Data Transmission Model}
We consider the decode-and-forward technique where at each hop the received message is decoded and re-encoded, if it is correctly decoded. Therefore, the message is successfully received by the destination if it is correctly decoded in all hops. Otherwise, outage occurs.  As the most promising HARQ approach leading to highest throughput/lowest outage probability \cite{throughputdef,MIMOARQkhodemun,a01661837,tuninetti2011}, we consider the incremental redundancy (INR) HARQ with a maximum of $M_i$ retransmissions in the $i$-th, $i=1,\ldots,T^\text{total},$ hop. Using INR HARQ with a maximum of  $M_i$ retransmissions, $q_i$ information nats are encoded into a \emph{parent} codeword of length $M_iL$ channel uses. The parent codeword is then divided into $M_i$ sub-codewords of length $L$ channel uses which are sent in the successive transmission rounds. Thus, the equivalent data rate, i.e., the code rate, at the end of round $m$ is $\frac{q_i}{mL}=\frac{R_i}{m}$ npcu where $R_i=\frac{q_i}{L}$ denotes the initial code rate in the $i$-th hop. In each round, the receiver combines all received sub-codewords to decode the message. Also, different independent channel realizations may be experienced in each round of HARQ. The retransmission continues until the message is correctly decoded or the maximum permitted transmission round is reached. Note that setting $M_i=1,\forall  i,$ represents the cases without HARQ, i.e., open-loop communication.

\section{Analytical results}
Consider the decode-and-forward approach in a multi-hop network consisting of $T$ RF- and $\tilde T$ FSO-based hops. Then, because independent channel realizations are experienced in different hops, the system outage probability is given by
\begin{align}\label{eq:Eqtotaloutage}
\Pr\left(\text{Outage}\right)=1-\prod_{i=1}^{T}{(1-\phi_i)}\prod_{i=1}^{\tilde T}{(1-\tilde \phi_i)},
\end{align}
where $\phi_i$ and $\tilde \phi_i$ denote the outage probability in the $i$-th RF- and FSO-based hops, respectively. Note that the order of the FSO and RF links therefore do not matter. To analyze the outage probability, we need to determine  $\phi_i$ and $\tilde \phi_i,\forall i$. Following the same procedure as in, e.g.,  \cite{throughputdef,MIMOARQkhodemun,a01661837} and using the properties of the imperfect PAs (\ref{eq:ampmodeldaniel}), the outage probability of the $i$-th RF- and FSO-based hops are found as
\begin{align}\label{eq:EqtotaloutageRF}
\phi_i=\Pr\Bigg(&\frac{1}{M_iC_i}\sum_{m=1}^{M_i}\sum_{c=(m-1)C_i+1}^{mC_i}\log\Bigg(1+\sqrt[1-\vartheta_i]{\frac{\epsilon_i P_i^{\text{cons}}}{(P_i^{\text{max}})^\vartheta_i}}G_i(c)\Bigg)\le\frac{R_i}{M_i}\Bigg), i=1,\ldots,T,
\end{align}
and
\begin{align}\label{eq:EqtotaloutageFSO}
&\tilde \phi_i=\Pr\Bigg(\frac{1}{M_i\tilde C_i}\sum_{m=1}^{M_i}\sum_{c=(m-1)\tilde C_i+1}^{m\tilde C_i}\log\left(1+\tilde P_i\tilde G_i(c)\right)\le\frac{R_i}{M_i}\Bigg),i=1,\ldots, \tilde T,
\end{align}
respectively. Here, (\ref{eq:EqtotaloutageRF})-(\ref{eq:EqtotaloutageFSO}) come from the maximum achievable rates of Gaussian channels where, in harmony with, e.g., \cite{6512100,7127443,6168189,5342330,5427418,throughputdef,7445896}, we have used the Shannon's capacity formula. Thus, our results provide a lower bound of outage probability which is tight for moderate/large codewords lengths. We assume that the FSO system is well-modeled as an additive white Gaussian noise channel, with insignificant signal-dependent shot noise contribution. Moreover, $\tilde P_i$ denotes the transmission power in the $i$-th, $i=1,\ldots,\tilde T,$ FSO-based  hop. We have considered a heterodyne detection technique in (\ref{eq:EqtotaloutageFSO}).  Also, with no loss of generality, we have normalized the receivers' noise variances. Hence, $P_i, \tilde P_i$ (in dB, $10\log_{10} P_i, 10\log_{10}\tilde P_i$) represent the SNR as well. Then, $C_i, i=1,\ldots,T,$ and $\tilde C_i, i=1,\ldots,\tilde T,$ represent the number of channel realizations experienced in each HARQ-based transmission round of the $i$-th RF- and FSO-based hops, respectively\footnote{ For simplicity, we present the results for cases with normalized symbol rates. However, using the same approach as in \cite{5342330}, it is straightforward to represent the results with different symbol rates of the links}. The number of channel realizations experienced within a codeword transmission is determined by the channel coherence times of the links, the codewords lengths, considered frequency as well as if diversity gaining techniques such as frequency hopping are utilized. Finally, $G_i(c)$ and $\tilde G_i(c)$ are the sum channel gains for the channel fading realization $c$ in the $i$-th RF- and FSO-based hop, respectively.

In the following, we present near-closed-form expressions for (\ref{eq:EqtotaloutageRF})-(\ref{eq:EqtotaloutageFSO}), and, consequently, (\ref{eq:Eqtotaloutage}). Then, Corollary 2 determines the ergodic achievable rate of multi-hop networks as well as the minimum number of required antennas in the RF-based hops to guarantee the ergodic achievable rate. Finally, Section III.C extends the results to mesh network.

Since there is no closed-form expression for the outage probabilities, we need to use different approximation techniques (see Table I for a summary of developed approximation schemes). In the first method, we concentrate on cases with long codewords where multiple channel realizations are experienced during data transmission in each hop, i.e., $C_i$ and $\tilde C_i,\forall i,$ are assumed to be large. Here, we use the central limit theorem (CLT) to approximate the contribution of the RF- and FSO-based hops by equivalent Gaussian random variables. Using the CLT, we find  different approximation results for the network outage probability/ergodic rate (Lemmas 1-5, Corollary 2). Then, Section III.B studies the system performance in cases with short codewords, i.e., small values of $C_i,\tilde C_i,\forall i$ (Lemma 6). It is important to note that the difference between the analytical schemes of Sections III.A and B comes from the values of the products $M_iC_i$ and $M_i\tilde C_i,\forall i.$ Therefore, from (\ref{eq:EqtotaloutageRF})-(\ref{eq:EqtotaloutageFSO}), the long-codeword results of Section III.A can be also mapped to cases with short codewords, large number of retransmissions and scaled code rates. As shown in Section IV, our derived analytical results are in {high agreement} with the numerical simulations.

\begin{table}\caption{Summary of developed approximation techniques.}\vspace{-2mm}
\center
\begin{tabular}{ | c | c | c | c |}
   \hline
    & Hop type & Metric & Tightness condition \\ \hline
   Lemma 1& RF & Outage probability & Long codeword, low SNR \\ \hline
   Lemmas 2-4& RF & Outage probability & Long codeword\\ \hline
   Lemma 5& FSO & Outage probability & Long codeword \\ \hline
   Corollary 1& RF, FSO & Outage probability & Long codeword \\ \hline
   Corollary 2& RF, FSO & Ergodic rate/required number of antennas & Long codeword \\ \hline
   Lemma 6& RF, FSO & Outage probability & Short codeword \\ \hline
\end{tabular}\vspace{-3mm}
\end{table}
\subsection{Performance Analysis with Long Codewords}
\textbf{\emph{Lemma 1:}} At low SNRs, the outage probability (\ref{eq:EqtotaloutageRF}) is approximately given by (\ref{eq:eqlemma2aa}) with $\mu_i$ and $\sigma_i^2$ defined in (\ref{eq:eqlowSNRlemma21}) and (\ref{eq:eqlowSNRlemma211}), respectively.
\begin{proof}
Using $\log(1+x)\simeq x$ for small values of $x$, (\ref{eq:EqtotaloutageRF}) is rephrased as
\begin{align}\label{eq:EqtotaloutageRFlow}
&\phi_i\simeq\Pr\Bigg(\frac{1}{M_iC_i}\sum_{m=1}^{M_i}\sum_{c=(m-1)C_i+1}^{mC_i}G_i(c)\le\frac{R_i}{M_i \sqrt[1-\vartheta_i]{\frac{\epsilon_i P_i^{\text{cons}}}{(P_i^{\text{max}})^\vartheta_i}}}\Bigg),
\end{align}
where for long codewords/large number of retransmissions, we can use the CLT to replace the random variable $\frac{1}{M_iC_i}\sum_{m=1}^{M_i}\sum_{c=(m-1)C_i+1}^{mC_i}G_i(c)$ by an equivalent Gaussian variable $\mathcal{V}_i\sim\mathcal{N}(\mu_i,\frac{1}{M_iC_i}\sigma_i^2)$ with
\begin{align}\label{eq:eqlowSNRlemma21}
&\mu_i=\int_0^\infty{xf_{G_i}(x)}\mathop  = \limits^{(a)}\frac{\Omega_i e^{-K_iN_i}N_i}{K_i+1}\prescript{}{1}F_{1}(N_i+1;N_i;K_iN_i),
\end{align}
and
\begin{align}\label{eq:eqlowSNRlemma211}
&
\sigma_i^2=\gamma_i-\mu_i^2,\nonumber\\&
\gamma_i=\int_0^\infty{x^2f_{G_i}(x)}\mathop  = \limits^{(b)}\frac{\Omega_i^2 e^{-K_iN_i(N_i+1)}N_i}{(K_i+1)^2}\prescript{}{1}F_{1}(N_i+2;N_i;K_iN_i).
\end{align}
Here, $\prescript{}{s}F_{t}(a_1,\ldots, a_s; b_1,\ldots,b_t;x)=\sum_{j=0}^\infty{\frac{(a_1)_j\ldots (a_s)_j}{(b_1)_j\ldots (b_t)_j}}\left(\frac{x^j}{j!}\right), (a)_0=1, (a)_j=a(a+1)\ldots(a+j-1),j>0,$ denotes the generalized hypergeometric function. Also, to find $(a)$-$(b)$ we have first used the property \cite[eq. (03.02.26.0002.01)]{wolframwebsite}
\begin{align}
I_n(x)=\frac{1}{\Gamma(n+1)}\left(\frac{x}{2}\right)^n \prescript{}{0}F_{1}\left(n+1;\frac{x^2}{4}\right),
\end{align}
to represent the PDF (\ref{eq:eqRiciansumpdf}) as
\begin{align}\label{eq:eqRiciansumpdf2}
f_{G_i}(x)&=\frac{(K_i+1)^{N_i}e^{-K_iN_i}}{\Omega_i^{N_i}\Gamma(N_i)}x^{N_i-1}e^{-\frac{(K_i+1)x}{\Omega_i}} \prescript{}{0}F_{1}\left(N_i;{\frac{K_i(K_i+1)N_ix}{\Omega_i}}\right),
\end{align}
and then derived $(a)$-$(b)$ based on the following integral identity \cite[eq. (7.522.5)]{bookhypergeometric}
\begin{align}\label{eq:eqinthypergeometric}
&\int_0^\infty{e^{-x}x^{\nu-1}}\prescript{}{s}F_{t}(a_1,\ldots,a_s;b_1,\ldots,b_t;\alpha x)\text{d}x
\nonumber\\&=\Gamma(\nu)\prescript{}{s+1}F_{t}(\nu,a_1,\ldots,a_s;b_1,\ldots,b_t;\alpha).
\end{align}
In this way, using the CDF of Gaussian random variables and the error function $\text{erf}(x)=\frac{2}{\sqrt{\pi}}\int_0^x{e^{-t^2}\text{d}t}$, (\ref{eq:EqtotaloutageRF}) is obtained by
\begin{align}\label{eq:eqlemma2aa}
\phi_i&\simeq\Pr\left(\mathcal{V}_i\le \frac{R_i}{M_i \sqrt[1-\vartheta_i]{\frac{\epsilon_i P_i^{\text{cons}}}{(P_i^{\text{max}})^\vartheta_i}}}\right)
=\frac{1}{2}\left(1+\text{erf}\left(\frac{\sqrt{M_iC_i}\left(\frac{R_i}{M_i \sqrt[1-\vartheta_i]{\frac{\epsilon_i P_i^{\text{cons}}}{(P_i^{\text{max}})^\vartheta_i}}}-\mu_i\right)}{\sqrt{2\sigma_i^2}}\right)\right),
\end{align}
as stated in the lemma.
\end{proof}

To present the second approximation method for (\ref{eq:EqtotaloutageRF}), we first represent an approximate expression for the PDF of the sum channel gain $G_i,\forall i,$ as follows.

\textbf{\emph{Lemma 2:}} For moderate/large number of antennas, which is of interest in MMW communication,  the sum gain $G_i,\forall i,$ is approximated by an equivalent Gaussian random variable $\mathcal{Z}_i\sim\mathcal{N}(N_i\zeta_i,N_i\nu_i^2)$ with $\zeta_i=\mathcal{S}_i(2)$, $\nu_i^2=\mathcal{S}_i(4)-\mathcal{S}_i(2)^2$ and $\mathcal{S}_i(n)\doteq\left(\frac{\Omega_i}{K_i+1}\right)^\frac{n}{2}\Gamma\left(1+\frac{n}{2}\right)\mathcal{L}_{\frac{n}{2}}\left(-K_i\right)$. Here, $\mathcal{L}_{n}(x)=\frac{e^x}{n!}\frac{\mathrm{d}^n }{\mathrm{d} x^n}\left(e^{-x}x^n\right)$ denotes the Laguerre polynomial of the $n$-th order and $K_i,\Omega_i$ are the fading parameters as defined in (\ref{eq:eqRicianpdf}).

\begin{proof}
Using the CLT for moderate/large number of antennas, the random variable  $G_i=\sum_{j_i=1}^{N_i}{g_i^{j_i}}$ is approximated by the Gaussian random variable $\mathcal{Z}_i\sim\mathcal{N}(N_i\zeta_i,N_i\nu_i^2)$. Here, from (\ref{eq:eqRicianpdf}), $\zeta_i$ and  $\nu_i^2$ are, respectively, determined by
\begin{align}\label{eq:eqmeanCLTmmw}
&\zeta_i=\int_0^\infty{xf_{{g_i^{j_i}}}(x)\text{d}x}=\frac{(K_i+1)e^{-K_i}}{\Omega_i}\int_0^\infty{xe^{-\frac{(K_i+1)x}{\Omega_i}}I_0\left(2\sqrt{\frac{K_i(K_i+1)x}{\Omega_i}} \right )\text{d}x},
\end{align}
and
\begin{align}\label{eq:eqmeanCLTmmw}
&\nu_i^2=\rho_i-\zeta_i^2,\nonumber\\&
\rho_i=\int_0^\infty{x^2f_{{g_i^{j_i}}}(x)\text{d}x}=\frac{(K_i+1)e^{-K_i}}{\Omega_i}\int_0^\infty{x^2e^{-\frac{(K_i+1)x}{\Omega_i}}I_0\left(2\sqrt{\frac{K_i(K_i+1)x}{\Omega_i}} \right )\text{d}x},
\end{align}
which, using the variable transform $t=\sqrt{x}$, some manipulations and the properties of the Bessel function $\frac{1}{b^2}\int_0^\infty{x^{n+1}e^{-\frac{x^2+c^2}{2b^2}}I_0\left(\frac{cx}{b^2}\right)\text{d}x}=b^n2^\frac{n}{2}\Gamma\left(1+\frac{n}{2}\right)\mathcal{L}\left(\frac{-c^2}{2b^2}\right),\forall c,b,n$, are determined as stated in the lemma.
\end{proof}

\textbf{\emph{Lemma 3:}} The outage probability (\ref{eq:EqtotaloutageRF}) is approximated by (\ref{eq:eqlemma3aa}) with $\hat\mu_i$ and $\hat\sigma_i^2$ given in (\ref{eq:eqtheoremmu2})-(\ref{eq:eqtheoremsigma2}), respectively.

\begin{proof}
Replacing the random variable $\frac{1}{M_iC_i}\sum_{m=1}^{M_i}\sum_{c=(m-1)C_i+1}^{mC_i}\log\left(1+\sqrt[1-\vartheta_i]{\frac{\epsilon_i P_i^{\text{cons}}}{(P_i^{\text{max}})^\vartheta_i}}G_i(c)\right)$ by its equivalent Gaussian random variable $\mathcal{U}_i\sim\mathcal{N}\left(\hat\mu_i,\frac{1}{M_iC_i}\hat\sigma_i^2\right)$,  the probability (\ref{eq:EqtotaloutageRF}) is rephrased as
\begin{align}\label{eq:eqapproxlemma3}
\phi_i\simeq\Pr\left(\mathcal{U}_i\le\frac{R_i}{M_i}\right), \mathcal{U}_i\sim\mathcal{N}\left(\hat\mu_i,\frac{1}{M_iC_i}\hat\sigma_i^2\right),
\end{align}
where
\begin{align}\label{eq:eqtheoremmu2}
\hat \mu_i&=\int_0^\infty{\log\left(1+\sqrt[1-\vartheta_i]{\frac{\epsilon_i P_i^{\text{cons}}}{(P_i^{\text{max}})^\vartheta_i}}x\right)}f_{G_i}(x)\text{d}x
\mathop  \simeq \limits^{(c)}
\int_0^\infty{\mathcal{Y}_i(x)f_{\mathcal{Z}_i}(x)\text{d}x}
\nonumber\\&=
\mathcal{Q}\left(\sqrt[1-\vartheta_i]{\frac{\epsilon_i P_i^{\text{cons}}}{(P_i^{\text{max}})^\vartheta_i}},0,N_i\zeta_i,N_i\nu_i^2,s_i\right)-\mathcal{Q}\left(\sqrt[1-\vartheta_i]{\frac{\epsilon_i P_i^{\text{cons}}}{(P_i^{\text{max}})^\vartheta_i}},0,N_i\zeta_i,N_i\nu_i^2,0\right)\nonumber\\&+\mathcal{Q}\left(r_i,\theta-r_id_i,N_i\zeta_i,N_i\nu_i^2,\infty\right)
-\mathcal{Q}\left(r_i,\theta-r_id_i,N_i\zeta_i,N_i\nu_i^2,s_i\right),
\nonumber\\&
\mathcal{Q}(a_1,a_2,a_3,a_4,x)\doteq-\frac{a_1a_3+a_2}{2}\text{erf}\left(\frac{a_3-x}{\sqrt{2a_4}}\right)-\frac{a_4}{2\pi}a_1e^{-\frac{(a_3-x)^2}{2a_4}},
\end{align}
and
\begin{align}\label{eq:eqtheoremsigma2}
&\hat\sigma_i^2=\hat\gamma_i-\hat\mu_i^2,\nonumber\\&
\hat \gamma_i=\int_0^\infty{\log^2\left(1+\sqrt[1-\vartheta_i]{\frac{\epsilon_i P_i^{\text{cons}}}{(P_i^{\text{max}})^\vartheta_i}}x\right)}f_{G_i}(x)\text{d}x
\mathop  \simeq \limits^{(d)}
\int_0^\infty{\mathcal{Y}_i^2(x)f_{\mathcal{Z}_i}(x)\text{d}x}
\nonumber\\&= \mathcal{T}\left(\sqrt[1-\vartheta_i]{\frac{\epsilon_i P_i^{\text{cons}}}{(P_i^{\text{max}})^\vartheta_i}},0,N_i\zeta_i,N_i\nu_i^2,s_i\right)-\mathcal{T}\left(\sqrt[1-\vartheta_i]{\frac{\epsilon_i P_i^{\text{cons}}}{(P_i^{\text{max}})^\vartheta_i}},0,N_i\zeta_i,N_i\nu_i^2,0\right)\nonumber\\&+\mathcal{T}\left(r_i,\theta-r_id_i,N_i\zeta_i,N_i\nu_i^2,\infty\right)
-\mathcal{T}\left(r_i,\theta-r_id_i,N_i\zeta_i,N_i\nu_i^2,s_i\right),
\nonumber\\&
\mathcal{T}\left(a_1,a_2,a_3,a_4,x\right)\doteq\frac{1}{2\sqrt{2\pi}}e^{-\frac{x^2+a_3^2}{2a_4}}\Bigg(\text{erf}\left(\frac{x-a_3}{\sqrt{2a_4}}\right)-2\sqrt{a_4}a_1e^{\frac{a_3x}{a_4}}(a_1(a_3+x)+2a_2)\nonumber\\&+\sqrt{2\pi}e^{\frac{x^2+a_3^2}{2a_4}}\left(a_1^2(a_3^2+a_4)+2a_1a_2a_3+a_2^2\right)
\Bigg).
\end{align}
Here, $(c)$ and $(d)$ in (\ref{eq:eqtheoremmu2}) and (\ref{eq:eqtheoremsigma2}) come from approximating $f_{G_i}(x)$ by $f_{\mathcal{Z}_i}(x)$ defined in Lemma 2 and the approximation $\log\left(1+\sqrt[1-\vartheta_i]{\frac{\epsilon_i P_i^{\text{cons}}}{(P_i^{\text{max}})^\vartheta_i}}x\right)\simeq \mathcal{Y}_i(x)$ where
%
%
%

\begin{align}
&\mathcal{Y}_i(x)=\left\{\begin{matrix}
{\sqrt[1-\vartheta_i]{\frac{\epsilon_i P_i^{\text{cons}}}{(P_i^{\text{max}})^\vartheta_i}}}x,\,\,\,\,\,\,\,\,\,\,\,\,\,\,\,\,\,\,\,\,\,\,\,\,\,\,\,\,\,\,\,\,\,\,\,\,\,\,\,\, x\in\left[0,s_i\right]\,\,\,\,\,\,\,\,\,\,\,\,\,\,\,\,\,\,\,\,\,\,\,\,\,\,\,\,\,\,\,\,\,\,\\
\theta+r_i\left(x-d_i \right ), \,\,\,\,\,\,\,\,\,\,\,\,\,\,\,\,\,\,\,\,\,\,\,\,\,\,\,\,\,\,\,\,\,\,  x> s_i,\,\,\,\,\,\,\,\,\,\,\,\,\,\,\,\,\,\,\,\,\,\,\,\,\,\,\,\,\,\,\,\,\,\,\,\,\,\,\,\,\,
\end{matrix}\right.,
\nonumber\\&
s_i=\frac{\theta }{\sqrt[1-\vartheta_i]{\frac{\epsilon_i P_i^{\text{cons}}}{(P_i^{\text{max}})^\vartheta_i}}\left(1-e^{-\theta}\right)}-\frac{1}{\sqrt[1-\vartheta_i]{\frac{\epsilon_i P_i^{\text{cons}}}{(P_i^{\text{max}})^\vartheta_i}}},
\nonumber\\&
r_i=\sqrt[1-\vartheta_i]{\frac{\epsilon_i P_i^{\text{cons}}}{(P_i^{\text{max}})^\vartheta_i}}e^{-\theta}
,
d_i=\frac{e^\theta-1}{\sqrt[1-\vartheta_i]{\frac{\epsilon_i P_i^{\text{cons}}}{(P_i^{\text{max}})^\vartheta_i}}}.
\end{align}
Then, following the same procedure as in (\ref{eq:eqlemma2aa}), (\ref{eq:eqapproxlemma3}) is obtained as
\begin{align}\label{eq:eqlemma3aa}
\phi_i&\simeq
\frac{1}{2}\left(1+\text{erf}\left(\frac{\sqrt{M_iC_i}\left(\frac{R_i}{M_i}-\hat\mu_i\right)}{\sqrt{2\hat\sigma_i^2}}\right)\right),\forall \theta>0.
\end{align}
Note that, in (\ref{eq:eqtheoremmu2})-(\ref{eq:eqlemma3aa}), $\theta>0$ is an arbitrary parameter and, based on our simulations, accurate approximations are obtained for a broad range of $\theta>0.$
\end{proof}
\textbf{\emph{Lemma 4:}} The outage probability of the RF-hop, i.e., (\ref{eq:EqtotaloutageRF}), is approximately given by
\begin{align}\label{eq:eqlemma3ss}
\phi_i&\simeq
\frac{1}{2}\left(1+\text{erf}\left(\frac{\sqrt{M_iC_i}\left(\frac{R_i}{M_i}-\breve{\mu}_i\right)}{\sqrt{2\breve{\sigma}_i^2}}\right)\right),
\end{align}
with $\breve{\mu}_i$ and $\breve{\sigma}_i^2$ defined in (\ref{eq:eqbrevemu}) and (\ref{eq:eqbrevesigma}), respectively.
\begin{proof}
  To prove the lemma, we again use the CLT where the achievable rate random variable $\frac{1}{M_iC_i}\sum_{m=1}^{M_i}\sum_{c=(m-1)C_i+1}^{mC_i}\log\left(1+\sqrt[1-\vartheta_i]{\frac{\epsilon_i P_i^{\text{cons}}}{(P_i^{\text{max}})^\vartheta_i}}G_i(c)\right)$ is replaced by  $\breve{\mathcal{U}}_i\sim\mathcal{N}(\breve{\mu}_i,\frac{1}{M_iC_i}\breve{\sigma}_i^2)$ with
\begin{align}\label{eq:eqbrevemu}
\breve{\mu}_i&
=\int_0^\infty{\log\left(1+\sqrt[1-\vartheta_i]{\frac{\epsilon_i P_i^{\text{cons}}}{(P_i^{\text{max}})^\vartheta_i}}x\right)f_{G_i}(x)\text{d}x}
\mathop  \simeq \limits^{(e)} \sqrt[1-\vartheta_i]{\frac{\epsilon_i P_i^{\text{cons}}}{(P_i^{\text{max}})^\vartheta_i}}\int_0^\infty{\frac{1-F_{\mathcal{Z}_i}(x)}{1+\sqrt[1-\vartheta_i]{\frac{\epsilon_i P_i^{\text{cons}}}{(P_i^{\text{max}})^\vartheta_i}}x}\text{d}x}
\nonumber\\&\mathop  \simeq \limits^{(f)} \sqrt[1-\vartheta_i]{\frac{\epsilon_i P_i^{\text{cons}}}{(P_i^{\text{max}})^\vartheta_i}}\int_0^\infty{\frac{\mathcal{W}_i(x)}{1+\sqrt[1-\vartheta_i]{\frac{\epsilon_i P_i^{\text{cons}}}{(P_i^{\text{max}})^\vartheta_i}}x}\text{d}x}
=\log\left(1+\sqrt[1-\vartheta_i]{\frac{\epsilon_i P_i^{\text{cons}}}{(P_i^{\text{max}})^\vartheta_i}}\left(\frac{-\sqrt{2\pi N_i\nu_i^2}}{2}+N_i\zeta_i\right)\right)\nonumber\\&+\mathcal{A}_i\Bigg(\sqrt[1-\vartheta_i]{\frac{\epsilon_i P_i^{\text{cons}}}{(P_i^{\text{max}})^\vartheta_i}},\frac{-1}{\sqrt{2\pi N_i\nu_i^2}},\frac{1}{2}+\frac{N_i\zeta_i}{\sqrt{2\pi N_i\nu_i^2}},\frac{-\sqrt{2\pi N_i\nu_i^2}}{2}+N_i\zeta_i\Bigg)\nonumber\\&-\mathcal{A}_i\Bigg(\sqrt[1-\vartheta_i]{\frac{\epsilon_i P_i^{\text{cons}}}{(P_i^{\text{max}})^\vartheta_i}},\frac{-1}{\sqrt{2\pi N_i\nu_i^2}},\frac{1}{2}+\frac{N_i\zeta_i}{\sqrt{2\pi N_i\nu_i^2}},\frac{\sqrt{2\pi N_i\nu_i^2}}{2}+N_i\zeta_i\Bigg),
\nonumber\\&
\mathcal{A}(a_1,a_2,a_3,x)\doteq\frac{a_1a_2x^2}{2}\log(1+a_1x)-\frac{a_1a_2x^2}{4}-\frac{a_2\log(1+a_1x)}{2a_1}-a_1a_3x+a_1a_3x\log(1+a_1x)\nonumber\\&+a_3\log(1+a_1x)+\frac{a_2x}{2}.
\end{align}
Here, $(e)$ comes from Lemma 2 and partial integration. Then, $(f)$ is obtained by the linearization technique $Q\left(\frac{x-N_i\zeta_i}{\sqrt{N_i\nu_i^2}}\right)\simeq \mathcal{W}_i(x)$ with

\begin{align}\label{eq:eqlinearappQ}
&
\mathcal{W}_i(x)\doteq\left\{\begin{matrix}
1 \,\,\,\,\,\,\,\,\,\text{if } x\le\frac{-\sqrt{2\pi N_i\nu_i^2}}{2}+N_i\zeta_i,\,\,\,\,\,\,\,\,\,\,\,\,\,\,\,\,\,\,\,\,\,\,\,\,\,\,\,\,\,\,\,\,\,\,\,\,\,\,\,\,\,\,\,\,\,\,\,\,\,\,\,\,\,\,\,\,\,\,\\
\frac{1}{2}-\frac{1}{\sqrt{2\pi N_i\nu_i^2}}(x-N_i\zeta_i) \,\,\,\,\,\,\,\,\,\,\,\,\,\,\,\,\,\,\,\,\, \,\,\,\,\,\,\,\,\,\,\,\,\,\,\,\,\,\,\,\,\,\,\,\,\,\,\,\,\,\,\,\,\,\,\,\,\,\,\,\,\,\,\,\,\,\,\,\,\,\,\,\,\,& \\\,\,\,\,\,\,\,\,\,\,\,\,\text{if } x\in\left[\frac{-\sqrt{2\pi N_i\nu_i^2}}{2}+N_i\zeta_i,\frac{\sqrt{2\pi N_i\nu_i^2}}{2}+N_i\zeta_i \right ],\,\,\,\,\,\\
0 \,\,\,\,\,\,\,\,\,\text{if } x>\frac{\sqrt{2\pi N_i\nu_i^2}}{2}+N_i\zeta_i,\,\,\,\,\,\,\,\,\,\,\,\,\,\,\,\,\,\,\,\,\,\,\,\,\,\,\,\,\,\,\,\,\,\,\,\,\,\,\,\,\,\,\,\,\,\,\,\,\,\,\,\,\,\,\,\,\,\,\,\,\,
\end{matrix}\right.
\end{align}
which is found by linearly approximating $Q\left(\frac{x-N_i\zeta_i}{\sqrt{N_i\nu_i^2}}\right)$ near the point $x=N_i\zeta_i.$ Finally, the last equality is obtained by partial integration and some manipulations. Also, following the same procedure, we have
\begin{align}\label{eq:eqbrevesigma}
&\breve{\sigma}_i^2=\breve{\rho}_i-\breve{\mu}_i^2\nonumber\\&
\breve{\rho}=\int_0^\infty{\log^2\left(1+\sqrt[1-\vartheta_i]{\frac{\epsilon_i P_i^{\text{cons}}}{(P_i^{\text{max}})^\vartheta_i}}x\right)f_{G_i}(x)\text{d}x}
\mathop  \simeq \limits^{} 2\sqrt[1-\vartheta_i]{\frac{\epsilon_i P_i^{\text{cons}}}{(P_i^{\text{max}})^\vartheta_i}}\int_0^\infty{\frac{\log\left(1+\sqrt[1-\vartheta_i]{\frac{\epsilon_i P_i^{\text{cons}}}{(P_i^{\text{max}})^\vartheta_i}}x\right)\mathcal{W}_i(x)}{1+\sqrt[1-\vartheta_i]{\frac{\epsilon_i P_i^{\text{cons}}}{(P_i^{\text{max}})^\vartheta_i}}x}\text{d}x}
\nonumber\\&
=\log^2\left(1+\sqrt[1-\vartheta_i]{\frac{\epsilon_i P_i^{\text{cons}}}{(P_i^{\text{max}})^\vartheta_i}}\left(\frac{-\sqrt{2\pi N_i\nu_i^2}}{2}+N_i\zeta_i\right)\right)\nonumber\\&+\mathcal{B}_i\Bigg(\sqrt[1-\vartheta_i]{\frac{\epsilon_i P_i^{\text{cons}}}{(P_i^{\text{max}})^\vartheta_i}},\frac{-1}{\sqrt{2\pi N_i\nu_i^2}},\frac{1}{2}+\frac{N_i\zeta_i}{\sqrt{2\pi N_i\nu_i^2}},\frac{-\sqrt{2\pi N_i\nu_i^2}}{2}+N_i\zeta_i\Bigg)\nonumber\\&-\mathcal{B}_i\Bigg(\sqrt[1-\vartheta_i]{\frac{\epsilon_i P_i^{\text{cons}}}{(P_i^{\text{max}})^\vartheta_i}},\frac{-1}{\sqrt{2\pi N_i\nu_i^2}},\frac{1}{2}+\frac{N_i\zeta_i}{\sqrt{2\pi N_i\nu_i^2}},\frac{\sqrt{2\pi N_i\nu_i^2}}{2}+N_i\zeta_i\Bigg),
\nonumber\\&
\mathcal{B}(a_1,a_2,a_3,x)\doteq\frac{a_1a_3-a_2}{a_1}\log^2(1+a_1x)-2a_2x+\frac{2a_1a_2x+2a_2}{a_1}\log(1+a_1x).
\end{align}

In this way, the outage probability is given by (\ref{eq:eqlemma3ss}).
\end{proof}
Finally, Lemma 5 represents the outage probability of the FSO-based hops as follows.

\textbf{\emph{Lemma 5:}} The outage probability of the FSO-based hop, i.e., (\ref{eq:EqtotaloutageFSO}), is approximately given by
\begin{align}\label{eq:eqlemma4aa}
\tilde\phi_i&\simeq
\frac{1}{2}\left(1+\text{erf}\left(\frac{\sqrt{M_i\tilde C_i}\left(\frac{R_i}{M_i}-\tilde\mu_i\right)}{\sqrt{2\tilde\sigma_i^2}}\right)\right),
\end{align}
where $\tilde \mu_i$ and $\tilde\sigma_i^2$ are given by (\ref{eq:mueq})-(\ref{eq:sigmaeq}) and \cite[eq. 43-44]{7445896} for the exponential and the Gamma-Gamma distributions of the FSO links, respectively.
\begin{proof}
Using the CLT, the random variable $\frac{1}{M_i\tilde C_i}\sum_{m=1}^{M_i}\sum_{c=(m-1)\tilde C_i+1}^{mC_i}\log\left(1+\tilde P_i\tilde G_i(c)\right)$ is approximated by its equivalent Gaussian random variable $\mathcal{R}_i\sim\mathcal{N}(\tilde\mu_i,\frac{1}{M_i\tilde C_i}\tilde\sigma_i^2)$,  where for the exponential distribution of the FSO link we have
\vspace{-0mm}
\begin{align}\label{eq:mueq}
\tilde \mu_i&=\int_0^\infty{f_{{\tilde G_i}}(x)\log\left(1+\tilde P_ix\right)\text{d}x}\mathop  = \limits^{(g)} \tilde P_i\int_0^\infty{\frac{1-F_{\tilde G_i}(x)}{1+\tilde P_ix}\text{d}x}=-e^{\frac{\tilde \lambda_i }{\tilde P_i}}\text{Ei}\left(-{\frac{\lambda_i }{\tilde P_i}}\right),
\end{align}
and
\vspace{-0mm}
\begin{align}\label{eq:sigmaeq}
&\tilde \sigma_i^2=\tilde \rho_i-\tilde \mu_i^2,\nonumber\\&
\tilde \rho_i=\int_0^\infty{f_{{\tilde  G_i}}(x)\log^2\left(1+\tilde P_ix\right)\text{d}x}  \mathop  = \limits^{(h)}{2\tilde P_i}\int_0^\infty{\frac{e^{- \lambda_i x}}{1+\tilde P_ix}\log\left(1+\tilde P_ix\right)\text{d}x}\mathop  = \limits^{(i)} \mathcal{H}_i\left(\infty\right)-\mathcal{H}_i\left(1\right),\nonumber\\&
\mathcal{H}_i(x)=2e^{\frac{\lambda_i}{\tilde P_i}}\bigg(\frac{\lambda_i}{\tilde P_i}x\prescript{}{3}F_{3}\left(1,1,1;2,2,2;-\frac{\lambda_i x}{\tilde P_i}\right)\nonumber\\&+\frac{1}{2}\log(x)\bigg(-2\left(\log\left(\frac{\lambda_i}{\tilde P_i}x\right)+\mathcal{E}\right)-2\Gamma\left(0,\frac{\lambda_i}{\tilde P_i}x\right)+\log(x)\bigg)\bigg).
\end{align}
Here, $\text{Ei}(x)=\int_x^\infty{\frac{e^{-t}\text{d}t}{t}}$ denotes the exponential integral function. Also,  $(g)$ and $(h)$ are obtained by partial integration. Then, denoting the Euler constant by $\mathcal{E}$, $(i)$ is given by the variable transformation $1+\tilde P_ix=t,$ some manipulations, as well as the definition of the Gamma incomplete function $\Gamma(s,x)=\int_x^\infty{t^{s-1}e^{-t}\text{d}t}$ and the generalized hypergeometric function $\prescript{}{a_1}F_{a_2}(\cdot).$ For the Gamma-Gamma distribution, on the other hand, the PDF $f_{\tilde G_i}$ in (\ref{eq:mueq})-(\ref{eq:sigmaeq}) is replaced by (\ref{eq:eqpdfgammagamma}) and the mean and variance are calculated by \cite[eq. 43]{7445896} and \cite[eq. 44]{7445896}, respectively. In this way, following the same arguments as in Lemmas 1, 3-4, the outage probability of the FSO-based hops is given by (\ref{eq:eqlemma4aa}).
\end{proof}
Lemmas 1-5 lead to different corollary statements about the performance of multi-hop RF-FSO systems, as stated in the following.

\textbf{\emph{Corollary 1:}} With long codewords, there are mappings between the performance of FSO- and RF-based hops, in the sense that the outage probability achieved in an RF-based hop is the same as the outage probability in an FSO-based hop experiencing specific long-term channel characteristics.
\begin{proof}
The proof comes from Lemmas 1-5 where for different hops the outage probability is given by the CDF of Gaussian random variables. Thus, with appropriate {long-term channel characteristics,} $(\mu_i,\sigma_i)$, $(\hat\mu_i,\hat\sigma_i)$, $(\breve{\mu}_i,\breve{\sigma}_i)$ and $(\tilde \mu_i,\tilde \sigma_i)$ in Lemmas 1 and 3-5 {can be equal leading to the same outage probability in these hops}.
\end{proof}
In this way, the performance of RF-FSO based multi-hop/mesh networks can be mapped to  ones using only the RF- or the FSO-based communication.

\textbf{\emph{Corollary 2:}} With asymptotically long codewords,
\begin{itemize}
  \item[1)] the minimum number of transmit antennas in an RF-based hop, such that the same rate is supported in all hops, is found by the solution of
\begin{align}\label{eq:eqlemma5min}
&N_i'=\mathop {\arg }\limits_{x}\Bigg\{\log\left(1+\sqrt[1-\vartheta_i]{\frac{\epsilon_i P_i^{\text{cons}}}{(P_i^{\text{max}})^\vartheta_i}}\left(\frac{-\sqrt{2\pi x\nu_i^2}}{2}+x\zeta_i\right)\right)\nonumber\\&+\mathcal{A}_i\Bigg(\sqrt[1-\vartheta_i]{\frac{\epsilon_i P_i^{\text{cons}}}{(P_i^{\text{max}})^\vartheta_i}},\frac{-1}{\sqrt{2\pi x\nu_i^2}},\frac{1}{2}+\frac{x\zeta_i}{\sqrt{2\pi x\nu_i^2}},\frac{-\sqrt{2\pi x\nu_i^2}}{2}+x\zeta_i\Bigg)\nonumber\\&-\mathcal{A}_i\Bigg(\sqrt[1-\vartheta_i]{\frac{\epsilon_i P_i^{\text{cons}}}{(P_i^{\text{max}})^\vartheta_i}},\frac{-1}{\sqrt{2\pi x\nu_i^2}},\frac{1}{2}+\frac{x\zeta_i}{\sqrt{2\pi x\nu_i^2}},\frac{\sqrt{2\pi x\nu_i^2}}{2}+x\zeta_i\Bigg)=\mathop {\min }\limits_{\forall j=1,\ldots,\tilde T}\{\tilde \mu_j\}\Bigg\},\forall i,
\end{align}
which can be calculated numerically.
  \item[2)] Also, the ergodic achievable rate of the multi-hop network is approximately given by
  \begin{align}\label{eq:eqergodicrate}
  \bar C(T,\tilde T)=\min\left(\mathop {\min }\limits_{\forall j=1,\ldots, T}\{ \breve{\mu}_j\},\mathop {\min }\limits_{\forall j=1,\ldots,\tilde T}\{\tilde \mu_j\}\right),
  \end{align}
   with $\breve{\mu_i}$ defined in (\ref{eq:eqbrevemu}) and $\tilde\mu_i$ given by (\ref{eq:mueq}) and \cite[eq. 43]{7445896} for the exponential and Gamma-Gamma distributions of the FSO link, respectively.
\end{itemize}

\begin{proof}
With asymptotically long codewords, i.e., very large $C_i,\tilde C_i,$ the achievable rates in the RF- and FSO-based hops converge to their corresponding ergodic capacity, and there is no need for HARQ because the data is always correctly decoded if it is transmitted with rates less than or equal to the ergodic capacity. Denoting the expectation operation by $E\{\cdot\}$, the ergodic capacity of an FSO-hop is given by $\tilde\mu_i=E\{\log(1+\tilde P_i\tilde G_i)\}$ which is determined by (\ref{eq:mueq}) and \cite[eq. 43]{7445896} for the exponential and Gamma-Gamma distributions of the FSO-based hop, respectively. For the RF-based hop, on the other hand, the ergodic capacity is found as $E\left\{\log\left(1+\sqrt[1-\vartheta_i]{\frac{\epsilon_i P_i^{\text{cons}}}{(P_i^{\text{max}})^\vartheta_i}}G_i\right)\right\}\simeq \breve{\mu}_i$ with $\breve{\mu}_i$ given in (\ref{eq:eqbrevemu}). In this way, the maximum achievable rate of the FSO-based hops is $\tilde R=\mathop {\min }\limits_{\forall j=1,\ldots,\tilde T}\{\tilde \mu_j\}$. Also, the minimum number of required antennas in the $i$-th RF-based hop is found by solving $\breve{\mu}_i=\tilde R$ which, from (\ref{eq:eqbrevemu}), leads to (\ref{eq:eqlemma5min}). Note that (\ref{eq:eqlemma5min}) is a single-variable equation and can be effectively solved by different numerical techniques.

Finally, following the same argument, the ergodic achievable rate of the RF-FSO network is given by (\ref{eq:eqergodicrate}), i.e., the maximum rate in which the data is correctly decoded in all hops.

\end{proof}

\subsection{Performance Analysis with Short Codewords}
Up to now, we considered the long-codeword scenario such that the CLT provides accurate approximation for
the sum of independent and identically distributed (IID) random variables. However, it is interesting to analyze the system performance in cases with short codewords, i.e., when $C_i$ and $\tilde C_i,\forall i,$ are small. This case is especially important for FSO links since the coherence time can be quite long (milliseconds). Here, we mainly concentrate on the Gamma-Gamma distribution
of the FSO-based hops. The same results as in \cite{5357980} can be applied to derive the outage probability of the FSO-based hops in the cases with exponential distribution.

\textbf{\emph{Lemma 6:}} For arbitrary numbers of $M_i, C_i$ and $\tilde C_i,$
\begin{itemize}
  \item[1)] The outage probabilities of the FSO- and RF-based hops are bounded by (\ref{eq:eqminkowseq1})
  and (\ref{eq:eqproofboundlemma6}), respectively.
  \item[2)] At low SNRs, a MISO-HARQ RF-based hop with $M_i$ retransmissions, $N_i$ transmit antennas and $C_i$ channel realizations per sub-codeword transmission can be mapped to an open-loop MISO setup with $M_iN_iC_i$ transmit antennas and single channel realization per codeword transmission.
\end{itemize}
\begin{proof}
Considering the FSO-based hops, one can use the Minkowski inequality \cite[Theorem 7.8.8]{minkowskibook}
\begin{align}\label{eq:eqminko}
\left(1+\left(\prod_{i=1}^n{x_i}\right)^\frac{1}{n}\right)^n\le \prod_{i=1}^{n}{\left(1+x_i\right)},
\end{align}
to write
\begin{align}\label{eq:eqminkowseq1}
&\tilde \phi_i=\Pr\left(\frac{1}{M_i\tilde C_i}\sum_{m=1}^{M_i}\sum_{c=(m-1)\tilde C_i+1}^{m\tilde C_i}{\log(1+\tilde P_i\tilde G_i(c))}\le \frac{R_i}{M_i}\right)\nonumber\\&= \Pr\left(\prod_{m=1}^{M_i}\prod_{c=(m-1)\tilde C_i+1}^{m\tilde C_i}\left(1+\tilde P_i\tilde G_i(c)\right)\le e^{\tilde C_iR_i}\right)\nonumber\\&\le \Pr\left(1+\tilde P_i\left(\prod_{m=1}^{M_i}\prod_{c=(m-1)\tilde C_i+1}^{m\tilde C_i}{\tilde G_i(c)}\right)^{\frac{1}{M_i\tilde C_i}}\le e^{\frac{R_i}{M_i}}\right)=F_{\mathcal{J}_i}\left(\left(\frac{e^{\frac{R_i}{M_i}}-1}{\tilde P_i}\right)^{M_i\tilde C_i}\right),
\end{align}
where, using the results of \cite[Lemma 3]{6168189} and for the Gamma-Gamma distribution of the variables $\tilde G_i$, the random variable $\mathcal{J}_i=\prod_{m=1}^{M_i}\prod_{c=(m-1)\tilde C_i+1}^{m\tilde C_i}{\tilde G_i(c)}$ follows the CDF
\begin{align}\label{eq:eqresminkows}
F_{\mathcal{J}_i}(x)&=\frac{1}{\Gamma^{M_i\tilde C_i}(a_i)\Gamma^{M_i\tilde C_i}(b_i)}\mathcal{G}_{1,2M_i\tilde C_i+1}^{2M_i\tilde C_i,1}\Bigg(({a_ib_i})^{M_i\tilde C_i}x\Bigg|_{\underbrace{a_i,a_i,\ldots,a_i}_{M_i\tilde C_i \text{ times}},\,\underbrace{b_i,b_i,\ldots,b_i}_{M_i\tilde C_i \text{ times}},0}^{\,\,\,\,\,\,\,\,1}\Bigg),
\end{align}
with $\mathcal{G}(.)$ denoting the Meijer G-function. Note that, the results of (\ref{eq:eqminkowseq1}) {are} mathematically applicable for every values of $M_i,\tilde C_i$. However, for, say $M_i\tilde C_i\ge 6$,
the implementation of the Meijer G-function in MATLAB is very time-consuming and the tightness of the  approximation decreases with $M_i,\tilde C_i$. Therefore, (\ref{eq:eqminkowseq1}) is useful for the performance analysis in
 cases with small $M_i,\tilde C_i,\forall i$, while the CLT-based approach of
Section III.A provides accurate performance evaluation for cases with long codewords.

For the RF-based hop, on the other hand, we use
\begin{align}\label{eq:eqauxi}
&\frac{1}{n}\log\left(1+\sum_{j=1}^n{x_j}\right)\le \frac{1}{n}\sum_{j=1}^{n}\log\left(1+x_j\right)\le\log\left(1+\frac{1}{n}\sum_{j=1}^n{x_j}\right),\forall n,x_j\ge 0,
\end{align}
to lower- and upper-bound the outage probability by
\begin{align}\label{eq:eqproofboundlemma6}
&\Pr\left(\log\left(1+\frac{\sqrt[1-\vartheta_i]{\frac{\epsilon_i P_i^{\text{cons}}}{(P_i^{\text{max}})^\vartheta_i}}}{M_iC_i}\sum_{m=1}^{M_i}\sum_{c=(m-1)C_i+1}^{mC_i}G_i(c)\right)\le\frac{R_i}{M_i}\right)
\nonumber\\&
\le \phi_i\le
\Pr\bigg(\frac{1}{M_iC_i}\log\left(1+{\sqrt[1-\vartheta_i]{\frac{\epsilon_i P_i^{\text{cons}}}{(P_i^{\text{max}})^\vartheta_i}}}\sum_{m=1}^{M_i}\sum_{c=(m-1)C_i+1}^{mC_i}G_i(c)\right)\le\frac{R_i}{M_i}\bigg)
\nonumber\\&\Rightarrow F_{\bold{G}_i}\left(\frac{M_iC_i\left(e^{\frac{R_i}{M_i}}-1\right)}{{\sqrt[1-\vartheta_i]{\frac{\epsilon_i P_i^{\text{cons}}}{(P_i^{\text{max}})^\vartheta_i}}}}\right)\le\phi_i\le F_{\bold{G}_i}\left(\frac{e^{{R_iC_i}}-1}{{\sqrt[1-\vartheta_i]{\frac{\epsilon_i P_i^{\text{cons}}}{(P_i^{\text{max}})^\vartheta_i}}}}\right).
\end{align}
Here, $\bold{G}_i=\sum_{m=1}^{M_i}\sum_{c=(m-1)C_i+1}^{mC_i}\sum_{j_i=1}^{N_i}g_i^{j_i}(c)$ is an equivalent sum channel gain variable with $M_iC_iN_i$ antennas at the transmitter whose PDF is obtained by replacing $N_i$ with $M_iC_iN_i$ in (\ref{eq:eqRiciansumpdf}). Also, $F_{\bold{G}_i}(\cdot)$ denotes the CDF of the equivalent sum channel gain variable.

To prove Lemma 6 part 2, we note that letting $x_i\to 0,\forall i,$ the inequalities in (\ref{eq:eqauxi}) are changed to equality. Thus, as a corollary result, at low SNRs a MISO-HARQ RF-link with $N_i$ transmit antennas, $M_i$ retransmissions and $C_i$ channel realizations within each retransmission round can be mapped to an open-loop MISO setup with $M_iC_iN_i$ transmit antennas, in the sense that the same outage probability is achieved in these setups.
\end{proof}
Note that the bounding schemes of (\ref{eq:eqproofboundlemma6}) are mathematically {applicable} for every values of $M_i,C_i.$ However, while the results of (\ref{eq:eqproofboundlemma6}) tightly match the exact numerical results for small values of $M_i,C_i$, the tightness decreases for large $M_i,C_i$'s. Thus, the results of Lemmas 1-4 and Lemma 6 can be effectively applied for the performance analysis of the RF-based hops in the cases with long and short codewords, respectively.  Finally, as another approximation for the cases with $M_i=1,C_i=1,$ we have
\begin{align}\label{eq:eqrffinite1}
\phi_{i|m_i=1,C_i=1}\simeq \frac{1}{2}\left(1+\text{erf}\left(\frac{\left(\frac{e^{R_i}-1}{ \sqrt[1-\vartheta_i]{\frac{\epsilon_i P_i^{\text{cons}}}{(P_i^{\text{max}})^\vartheta_i}}}-N_i\zeta_i\right)}{\sqrt{2N_i\nu_i^2}}\right)\right),
\end{align}
which comes from Lemma 2.


\subsection{Performance Analysis in Mesh Networks}
Consider a mesh network consisting of $\mathcal{X}$ non-overlapping routes from the source to the destination with independent channel realizations for the hops. The $\mathcal{x}$-th, $\mathcal{x}=1,\ldots,\mathcal{X},$ route is made of $T_\mathcal{x}$ RF- and $\tilde T_\mathcal{x}$ FSO-based hops and the routes can have different total number of hops $T_\mathcal{x}^\text{total}=T_\mathcal{x}+\tilde T_\mathcal{x},\mathcal{x}=1,\ldots,\mathcal{X}$. In this case, the network outage probability is given by
\begin{align}\label{eq:eqoutagemesh}
\Pr(\text{Outage})^\text{mesh}=\prod_{\mathcal{x}=1}^{\mathcal{X}}\left(\Pr\left(\text{Outage}_\mathcal{x}\right)\right),
\end{align}
where $\Pr(\text{Outage}_\mathcal{x})$ is the outage probability in the $\mathcal{x}$-th route as given in (\ref{eq:Eqtotaloutage}). In (\ref{eq:eqoutagemesh}), we have used the fact that in a mesh network an outage occurs if the data is correctly transferred to the destination through none of the routes. With the same arguments, the ergodic achievable rate of the mesh network is obtained by
\begin{align}\label{eq:eqrateagemesh}
\bar C^\text{mesh}=\mathop {\max }\limits_{\forall \mathcal{x}=1,\ldots,\mathcal{X}}\left\{\bar C_\mathcal{x}\right\},
\end{align}
with $\bar C_\mathcal{x}$ derived in (\ref{eq:eqergodicrate}). This is based on the fact that, knowing the long-term channel characteristics, one can set the data rate equal to the maximum achievable rate of the best route and the message is always correctly decoded by the destination, if the codewords are asymptotically long. The performance of mesh networks is studied in Fig. 9.

\section{{Numerical} Results}
Throughout the paper, we presented different approximation techniques. The verification of these results is demonstrated in Figs. 1, 2, 6-8 and, as seen in the sequel, the analytical results follow the {numerical} results with high accuracy. Then, to avoid too much information in each figure, Figs. 3-5, 9 report only the simulation results. Note that in all figures we have double-checked the results with the ones obtained analytically, and they match tightly.

The simulation results are presented for homogenous setups. That is, different RF-based hops follow the same long-term fading parameters $K_i,\omega_i,\forall i,$ in (\ref{eq:eqRicianpdf})-(\ref{eq:eqRiciansumpdf}), and the FSO-based hops also experience the same long-term channel parameters, i.e., $\lambda_i,a_i$ and $b_i$ in (\ref{eq:Eqexpfsopdf})-(\ref{eq:eqpdfgammagamma}). Moreover, we set $M_i=M_j$ and $R_i=R_j,\forall i,j=1,\ldots, T^\text{total}.$ In all figures, we set  $\tilde P_i=N_iP_i^\text{cons}$ such that the total consumed power at different hops is the same. Then, using (\ref{eq:ampmodeldaniel}), one can determine the output power of the RF-based antennas. Also, because the noise variances are set to 1, $\tilde P_i$ (in dB, $10\log_{10}\tilde P_i$) is referred to the SNR as well. In Figs. 1, 2 and 5, we assume an ideal PA. The effect of non-ideal PAs is verified in Figs. 3, 4, 6-9. With non-ideal PAs, we consider the state-of-the-art parameter settings $\vartheta_i=0.5, \epsilon_i=0.75, P_i^\text{max}=25 \text{ dB},\forall i,$ \cite{phdthesisBjornemo,6515206,6725577,7104158}, unless otherwise stated.

The parameters of the Rician RF PDF (\ref{eq:eqRicianpdf}) are set to $\omega_i=1, K_i=0.01, \forall i,$ leading to unit mean and variance of the channel gain distribution $f_{g_i^{j_i}}(x),\forall i,j_i$. With the exponential distribution of the FSO-based hops, we consider $f_{\tilde G_i}(x)=\lambda_i e^{-\lambda_i x}$ with $\lambda_i=1,\forall i.$ Also, for the Gamma-Gamma distribution we set $f_{\tilde G_i}(x)=\frac{2(a_ib_i)^{\frac{a_i+b_i}{2}}}{\Gamma(a_i)\Gamma(b_i)}x^{\frac{a_i+b_i}{2}-1}\mathcal{K}_{a_i-b_i}\left(2\sqrt{a_ib_ix}\right),$ $a_i=4.3939, b_i=2.5636, \forall i,$ which corresponds to Rytov variance of 1 \cite{6168189}. Figures 1-8 consider multi-hop networks. The performance of mesh networks is studied in Fig. 9. Note that, as discussed in Section III, the results of  cases with long codewords and few number of retransmissions can be mapped to the cases with short codewords, large number of retransmissions and a scaled code rate (see (\ref{eq:EqtotaloutageRF})-(\ref{eq:EqtotaloutageFSO})). Finally, it is worth noting that we have verified the analytical and the numerical results for a broad range of parameter settings, which, due to space limits and because they lead to the same qualitative conclusions as in the presented figures, are not reported in the figures.

\begin{figure}
\centering
  \includegraphics[width=0.6\columnwidth]{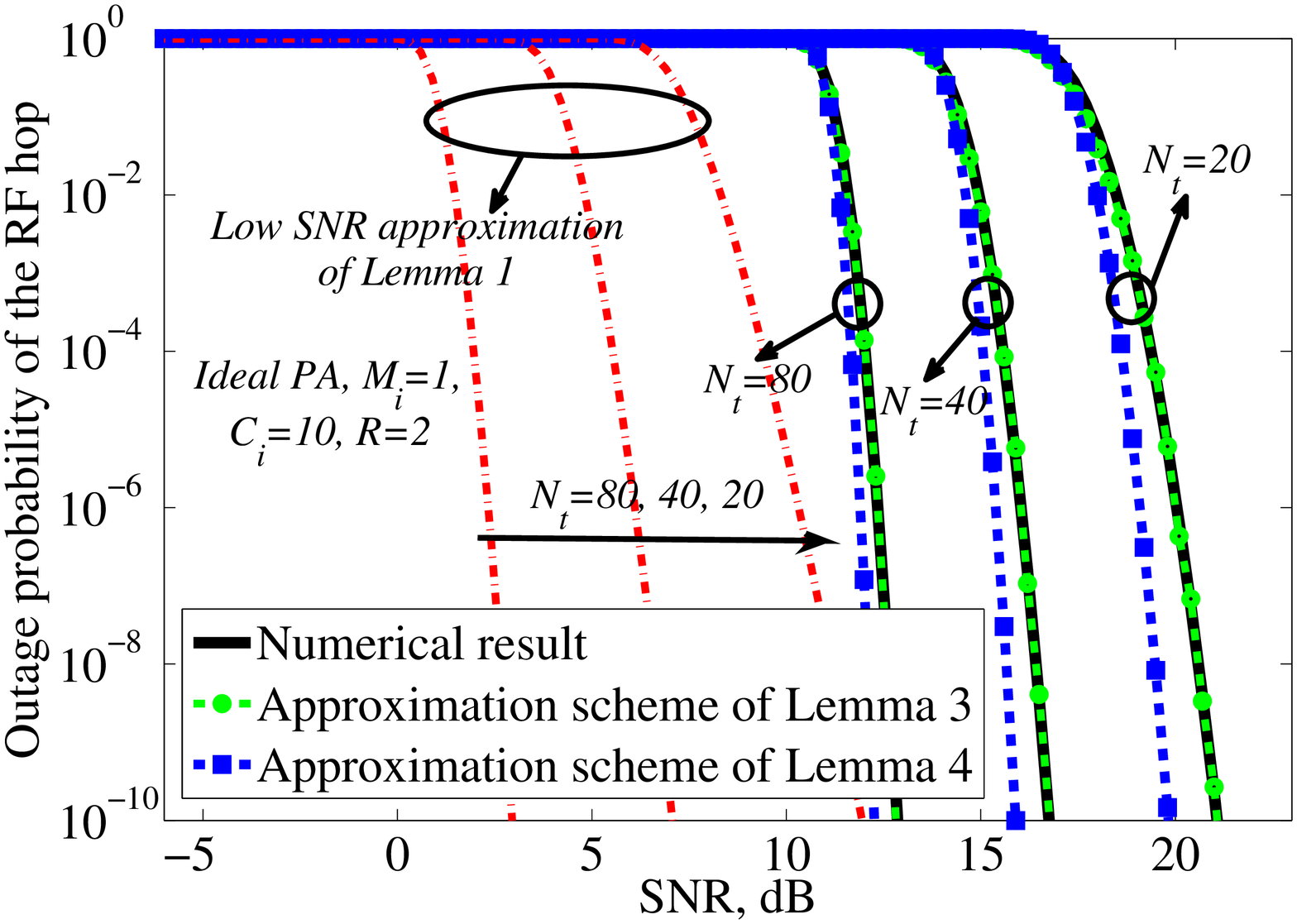}\\\vspace{-6mm}
\caption{On the tightness of the results in Lemmas 1-4. Ideal PA, single RF-based hop, $M_i=1, R=2, C_i=10.$ The results are presented for $N_i=20, 40,$ and $80,\forall i.$}\vspace{-6mm}\label{figure111}
\end{figure}

\begin{figure}
\centering
  \includegraphics[width=0.6\columnwidth]{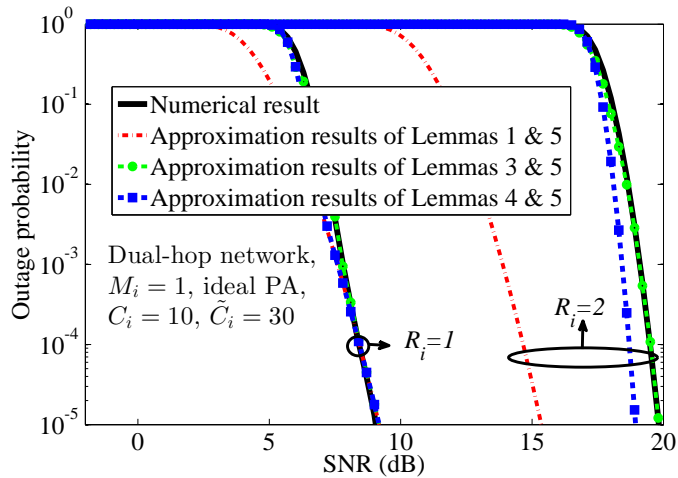}\\\vspace{-6mm}
\caption{On the tightness of the results in Lemmas 1-5. Ideal PA, dual-hop network, $M_i=1, R_i=1,2, C_i=10, \tilde C_i=30, T=1,\tilde T=1,$ and $N_i=20, \forall i.$}\vspace{-6mm}\label{figure111}
\end{figure}

\begin{figure}
\centering
  \includegraphics[width=0.6\columnwidth]{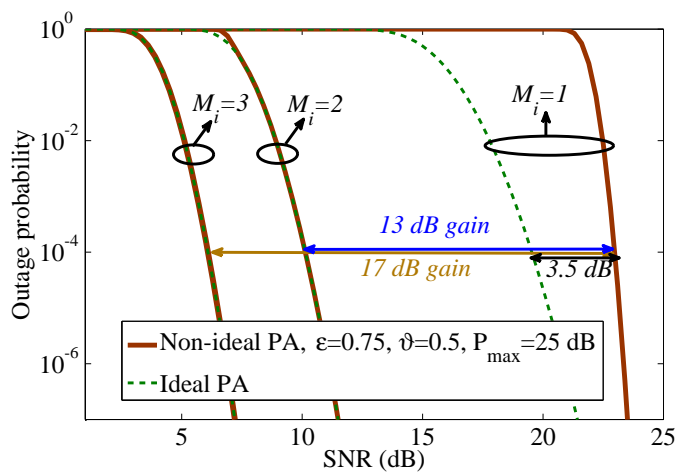}\\\vspace{-6mm}
\caption{Outage probability of a dual-hop RF-FSO network for different PA models and maximum number of retransmissions, $M_i,\forall i$. Exponential distribution of the FSO link, $T=1, \tilde T=1, C_i=10, \tilde C_i=20, R_i=3$ npcu, and  $N_i=60, \forall i.$}\vspace{-6mm}\label{figure111}
\end{figure}

\begin{figure}
\centering
  \includegraphics[width=0.6\columnwidth]{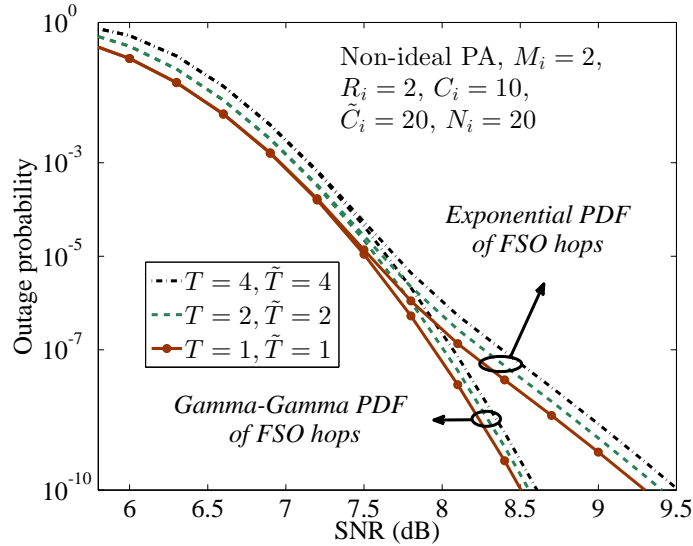}\\\vspace{-6mm}
\caption{The outage probability for different numbers of RF- and FSO-based hops, i.e., $T$ and $\tilde T$. Non-ideal PA, $\vartheta_i=0.5, \epsilon_i=0.75, P_i^\text{max}=25$ dB, $M_i=2, N_i=20, R_i=2$ npcu, $C_i=10,$ and  $\tilde C_i=20, \forall i.$}\vspace{-6mm}\label{figure111}
\end{figure}

\begin{figure}
\centering
  \includegraphics[width=0.6\columnwidth]{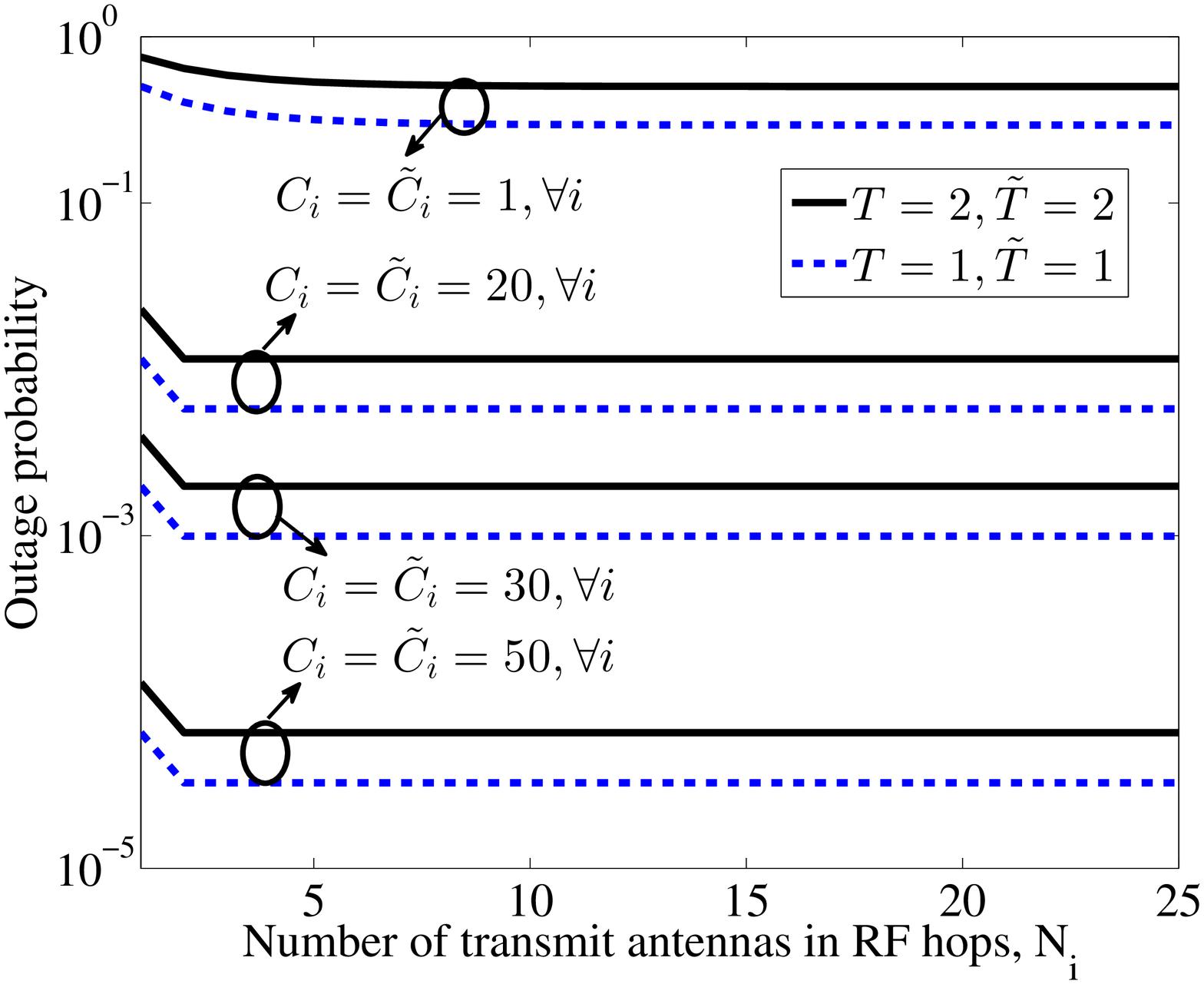}\\\vspace{-6mm}
\caption{Outage probability for different numbers of transmit antennas in the RF-based hops and channel coherence times. Exponential distribution of the FSO-based hops, ideal PA, $R_i=1.5$ npcu, $M_i=1, T_i=\tilde T_i=1, \forall i,$ and $\text{SNR}=10$ dB. }\vspace{-6mm}\label{figure111}
\end{figure}

The simulation results are presented in different parts as follows.

\emph{On the approximation approaches of Lemmas 1-5:} Considering an ideal PA, $M_i=1$ (as the worst-case scenario), $R_i=2$ npcu, and $C_i=10, \forall i,$ Fig. 1 verifies the tightness of the approximation schemes of Lemmas 1-4. Particularly, we plot the outage probability of an RF-based hop for different numbers of transmit antennas $N_i, \forall i.$ Then, Fig. 2 demonstrates the outage probability of a dual-hop RF-FSO setup versus the SNR.  Here, we set $M_i=1, R_i=1,2$ npcu, and $C_i=10, \tilde C_i=30, N_i=20, T=1, \tilde T=1 \forall i,$ and the results are presented for cases with ideal PAs at the RF-based hops. As  observed, the analytical results of Lemmas 2-5 mimic the exact results with very high accuracy (Figs. 1-2). Also, Lemma 1  properly approximates the outage probability at low and high SNRs, and the tightness increases as the code rate decreases (Fig. 2). Moreover, the tightness of the approximation results of Lemmas 3-4 increases with the number of RF-based transmit antennas (Fig. 1). This is because the tightness of the CLT-based approximations in Lemma 2 increases with $N_i,\forall i.$ Finally, although not demonstrated in Figs. 1-2, the tightness of the CLT-based approximation schemes of Lemmas 3-5 increases with the maximum number of retransmissions $M_i,\forall i$.

\emph{On the effect of HARQ and imperfect PAs:} Shown in Fig. 3 is the outage probability of a dual-hop RF-FSO network for different maximum numbers of HARQ-based retransmission rounds $M_i,\forall i.$ Also, the figure compares the system performance in cases with ideal and non-ideal PAs. Here, the results are obtained for the exponential distribution of the FSO link, $T=1, \tilde T=1, C_i=10, \tilde C_i=20, R_i=3$ npcu, and  $N_i=60, \forall i.$ As demonstrated, with no HARQ, the efficiency of the RF-based PAs affects the system performance considerably. For instance, with the parameter settings of the figure and outage probability $10^{-4},$ the PAs inefficiency increases the required power by $3.5$ dB. On the other hand, the HARQ can effectively compensate the effect of imperfect PAs, and the difference between the outage probability of the cases with ideal and non-ideal PAs is negligible for $M>1.$ Also, the effect of non-ideal PA decreases at high SNRs which is intuitively because the \emph{effective} efficiency of the PAs $\epsilon_i^\text{effective}=\epsilon_i(\frac{P_i}{P_i^\text{max}})^{\vartheta_i},\forall i,$ is improved as the SNR increases. Finally, the implementation of HARQ improves the energy efficiency significantly. As  an example, consider the outage probability $10^{-4}$, an ideal PA and the parameter settings of Fig. 3. Then, compared to the open-loop communication, i.e., $M_i=1$, the implementation of HARQ with a maximum of 2 and 3 retransmissions reduces the required power by 13 and 17 dB, respectively.

\begin{figure}
\centering
  \includegraphics[width=0.6\columnwidth]{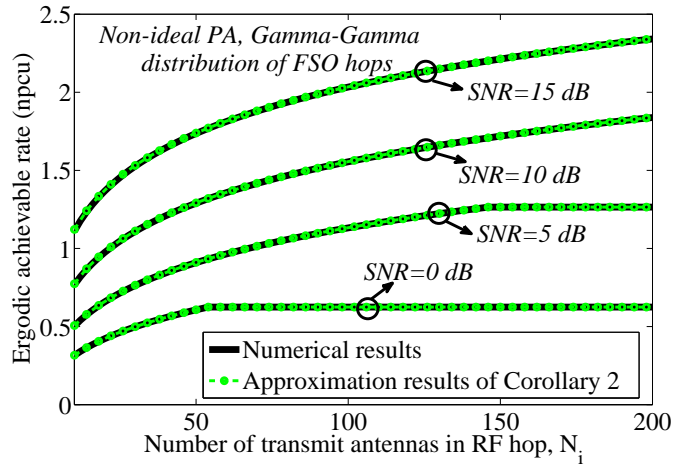}\\\vspace{-6mm}
\caption{Ergodic achievable rate for different numbers of transmit antennas in the RF-based hops and SNRs. Gamma-Gamma distribution of the FSO-based hops, non-ideal PA $\vartheta_i=0.5, \epsilon_i=0.75,$ and $P_i^\text{max}=25 \text{ dB},\forall i.$}\vspace{-6mm}\label{figure111}
\end{figure}

\begin{figure}
\centering
  \includegraphics[width=0.6\columnwidth]{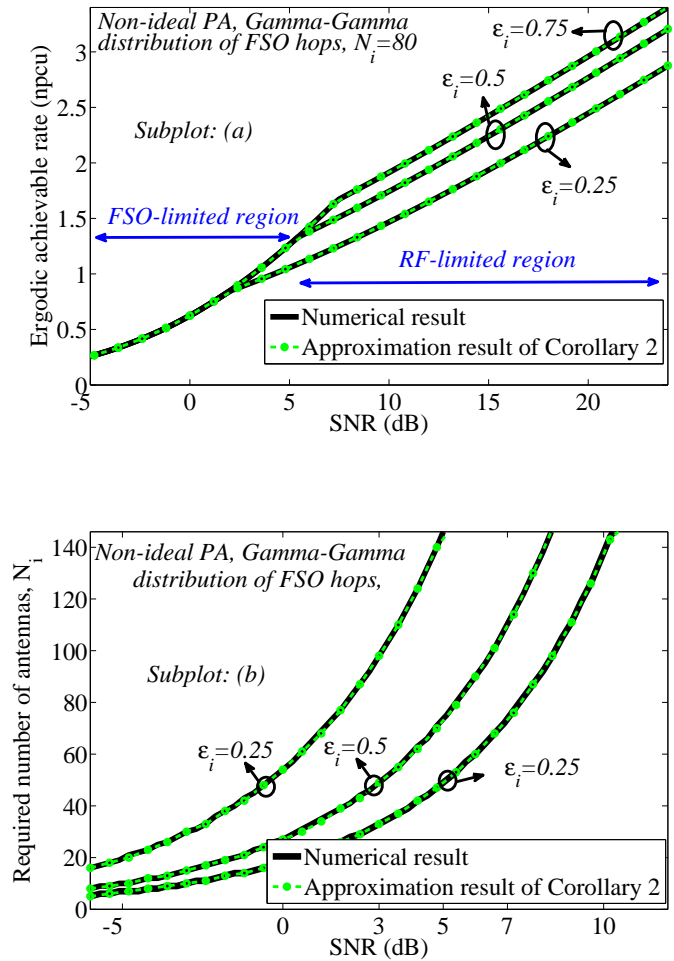}\\\vspace{-6mm}
\caption{On the tightness of the analytical results of Corollary 2. Subplot (a): Ergodic achievable rate vs the SNR. Non-ideal PA, Gamma-Gamma distribution of the FSO hops, $N_i=80, \forall i$. Subplot (b): The minimum number of transmit antennas in the RF-based hops to guarantee the same rate as in the FSO-based hops. Non-ideal PA and Gamma-Gamma distribution of the FSO hops. For the non-ideal PAs, we have $\vartheta_i=0.5,\epsilon_i=0.75,$ and  $P_i^\text{max}=25 \text{ dB},\forall i.$}\vspace{-6mm}\label{figure111}
\end{figure}

\begin{figure}
\centering
  \includegraphics[width=0.6\columnwidth]{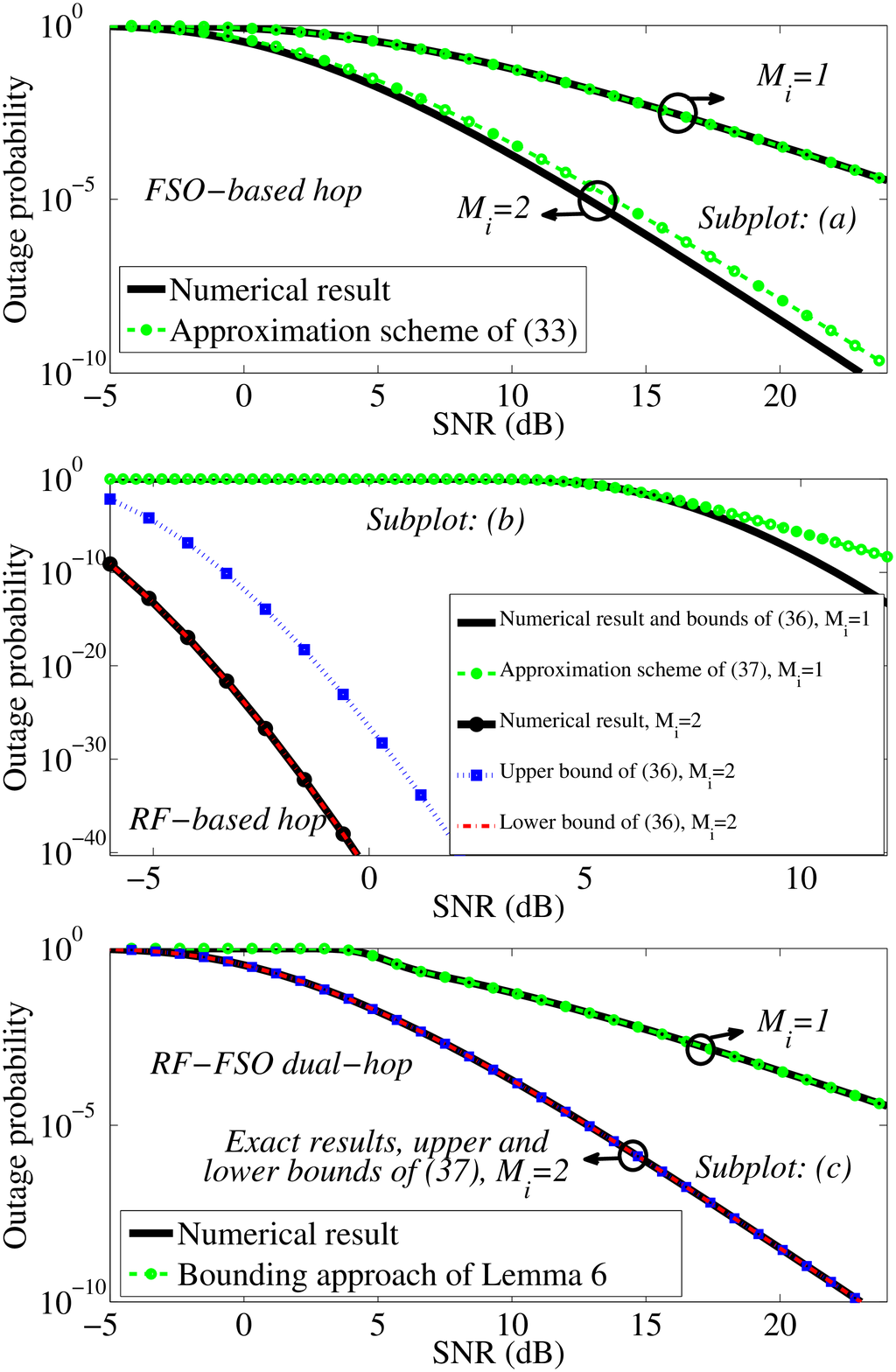}\\\vspace{-6mm}
\caption{Outage probability in the cases with short codewords. Non-ideal PA, Gamma-Gamma distribution of the FSO hops, $R_i=1$ npcu, $M_i=1, C_i=1,\tilde C_i=1, N_i=60, \forall i$. For the non-ideal PAs, we have $\vartheta_i=0.5,\epsilon_i=0.75,$ and  $P_i^\text{max}=25 \text{ dB},\forall i.$ In subplots a-c, the outage probability is presented for an FSO-based hop, an RF-based hop and a dual-hop RF-FSO network, respectively. }\vspace{-6mm}\label{figure111}
\end{figure}

\begin{figure}
\centering
  \includegraphics[width=0.6\columnwidth]{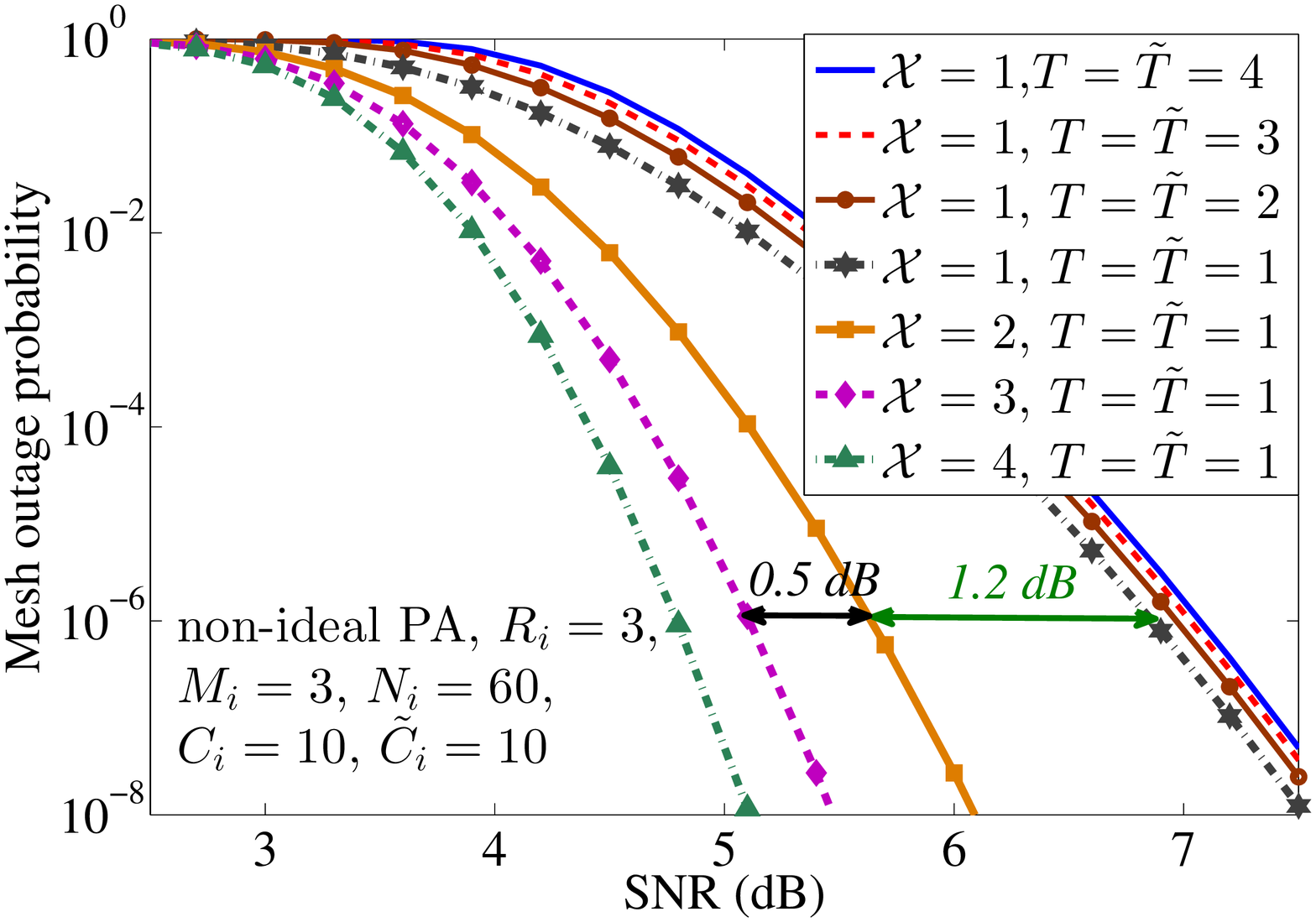}\\\vspace{-6mm}
\caption{Outage probability of a mesh network for different numbers of routes and hops. Non-ideal PA, Gamma-Gamma distribution of the FSO hops, $R_i=3$ npcu, $M_i=3, C_i=10,\tilde C_i=10,$ and $N_i=60, \forall i$. For the non-ideal PAs, we have $\vartheta_i=0.5,\epsilon_i=0.75,$ and  $P_i^\text{max}=25 \text{ dB},\forall i.$}\vspace{-6mm}\label{figure111}
\end{figure}

\emph{System performance with different numbers of hops:} In Fig. 4, we demonstrate the outage probability in cases with different numbers of RF- and FSO-based hops, i.e., $T, \tilde T$. In harmony with intuition, the outage probability increases with the number of hops. However, the outage probability increment is negligible, particularly at high SNRs, because the data is correctly decoded with high probability in different hops as the SNR increases. Finally, as a side result, the figure indicates that the outage probability of the RF-FSO based multi-hop network is not sensitive to the distribution of the FSO-based hops at low SNRs. This is intuitive because, at low SNRs and with the parameter settings of the figure, the outage event mostly occurs in the RF-based hops. However, at high SNRs where the outage probability of different hops are comparable, the PDF of the FSO-based hops affects the network performance.

\emph{On the effect of RF-based transmit antennas:} Considering an exponential distribution of the FSO-based hops, ideal PAs, $R_i=1.5$ npcu, $M_i=1, T_i=\tilde T_i=1, \forall i,$ and $\text{SNR}=10$ dB, Fig. 5 demonstrates the effect of the number of RF transmit  antennas on the network outage probability. Also, the figure compares the system performance in cases with short and long codewords, i.e., in cases with small and large values of $C_i,\tilde C_i.$ As seen, with short codewords, the outage probability decreases with the number of RF-based transmit antennas monotonically. This is because, with the parameter settings of the figure, the data is correctly decoded with higher probability as the number of antennas increases. With long codewords, on the other hand, the outage probability is (almost) insensitive to the number of transmit antennas for $N_i\ge 3$. Finally, the outage probability decreases with $C_i,\tilde C_i,$ because the HARQ exploits time diversity as more channel realizations are experienced within each codeword transmission.

In Fig. 6, we plot the network ergodic achievable rates for cases with Gamma-Gamma distribution of the FSO-based hops, non-ideal PAs and different numbers of transmit antennas/SNRs. Also, the figure verifies the accuracy of the approximation schemes of Corollary 2. Note that, due to the homogenous network structure, the ergodic rate is independent of the number of RF- and FSO-based hops. As seen, at low/moderate SNRs, the network ergodic rate increases (almost) logarithmically with the number of RF antennas. At high SNRs, on the other hand, the ergodic rate becomes independent of the number of RF transmit antennas. This is because with large number of RF-based antennas the achievable rate of the RF-based hops exceeds the one in FSO-based hops, and the network ergodic rate is given by the achievable rate of the FSO-based hops. Finally, the number of antennas above which the ergodic rate is limited by the achievable rate of the FSO-based hops increases with the SNR.

\emph{On the ergodic achievable rates:} Along with Fig. 6, we evaluate the accuracy of the results of Corollary 2 in Figs. 7a and 7b. Particularly, Fig. 7a demonstrates the network ergodic rate for different PA models and compares the simulation results with the ones derived in (\ref{eq:eqergodicrate}). Then, Fig. 7b verifies the accuracy of (\ref{eq:eqlemma5min}). Here, we show the minimum number of required RF transmit antennas versus the SNR which determines the ergodic rate of the FSO-based hops.   As can be seen, the approximation results of Corollary 2 are very tight for a {broad} range of parameter settings. Thus, (\ref{eq:eqlemma5min}) and (\ref{eq:eqergodicrate}) can be effectively used to derive the required number of RF transmit antennas and the network ergodic rate, respectively (Figs. 7a and 7b). The ergodic rate shows different behaviors in the, namely, FSO-limited and RF-limited regions. With the parameter settings of the figure, the ergodic rate is limited by the achievable rates of the FSO-based hops at low SNRs (FSO-limited region in Fig. 7a). However, as the SNR increases, the achievable rates of the FSO-based hops exceed the ones in the RF-based hops and, consequently, the network ergodic rate is limited by the rate of the RF-based hops (RF-limited region in Fig. 7a). As a result, the efficiency of the RF PAs affects the ergodic rate at high SNRs. Finally, the figure indicates that at high SNRs the network ergodic rate increases (almost) linearly with the SNR.

As shown in Fig. 7b, the required number of RF-based transmit antennas to guarantee the same rate as in the FSO-based hops increases considerably with the SNR and, consequently, the ergodic rate of the FSO-based hops. Moreover, the PAs efficiency affects the required number of antennas significantly. As an example, consider the parameter settings of Fig. 7b and $\text{SNR} = 3$ dB. Then, the required number of RF-based antennas is given by 33, 49 and 98 for the cases with PA efficiency $75\%$, $50\%$ and $25\%$, respectively. Thus, hardware impairments such as the PA inefficiency affect the system performance remarkably and should be carefully considered in the network design. However, selecting the proper number of antennas and PA properties is not easy because the decision depends on several parameters such as complexity, infrastructure size and cost.


\emph{Performance analysis with short codewords:} In Figs. 8a, 8b and 8c, we study the outage probability of an FSO-based hop, an RF-based hop and a dual-hop RF-FSO network, respectively. Particularly, considering non-ideal PAs and Gamma-Gamma distribution of the FSO-based hops, the results are obtained for $C_i=\tilde C_i=1,$ $M_i=1,2 \forall i,$ and we evaluate the accuracy of different bounds/approximations in Lemma 6 and (\ref{eq:eqrffinite1}). As demonstrated, the bound of (\ref{eq:eqminkowseq1}) matches the exact values derived via simulation analysis of $\tilde \phi_i$ exactly in cases with $M_i=1$. Also, the bounding/approximation methods of (\ref{eq:eqminkowseq1}), (\ref{eq:eqproofboundlemma6}) and (\ref{eq:eqrffinite1}) mimic the numerical results with high accuracy in cases with a maximum of $M_i=2, \forall i,$ retransmissions. Thus, the results of Section III.B can be efficiently used to analyze the RF-FSO systems in  cases with small values of $M_i, C_i,\tilde C_i,\forall i$.

\emph{On the performance of mesh networks:} In Fig. 9, we study the outage probability of mesh networks for different numbers of routes. Here, we consider non-ideal PAs, exponential PDF of the FSO hops, $M_i=3,$  $N_i=60,$ $C_i=10, \tilde C_i=10,$ and $R_i=3$ npcu. The results are presented for  cases with one RF- and one FSO-based hop in each route. Also, we compare the outage probability of the mesh network with {that of a} single route setup consisting of different numbers of RF- and FSO-based hops. Note that there are the same total number of RF- and FSO-based hops in each case with $\mathcal{X}=n,T_\mathcal{x}=\tilde T_\mathcal{x}=1,\forall \mathcal{x}=1,\ldots,\mathcal{X},$ and $\mathcal{X}=1,T_1=\tilde T_1=n$.

As demonstrated in Fig. 9, in contrast to the single-route setup where the outage probability increases with the number of hops, the outage probability of the mesh network decreases considerably by adding more parallel routes into the network. For instance, consider the parameter settings of the figure and outage probability of $10^{-6}.$ Then, compared to  cases with a single route, the required SNR at each hop decreases by almost $1.2$ dB if the data is transferred through two routes. This is intuitive because the probability that the data is {correctly} received by the destination increases with the number of routes. However, the relative effect of adding more routes decreases with the number of routes, and, for the  example parameters considered, there is about $0.5$ dB energy efficiency improvement if the number of routes increases from $\mathcal{X}=2$ to $\mathcal{X}=3.$ Finally, while we did not consider it in Section III.C, the performance of the mesh network is further improved if the signals from different routes are combined at the destination.

\section{Conclusion}
We studied the performance of RF-FSO based multi-hop and mesh networks in  cases with short and long codewords. Considering different channel conditions, we derived closed-form expressions for the networks outage probability, the ergodic rates as well as the required number of RF transmit antennas to guarantee different achievable rate quality-of-service requirements. The results are presented for  cases with and without HARQ. As demonstrated, depending on the codeword length, there are different methods for analytical performance evaluation of the multi-hop/mesh networks. Moreover, there are mappings between the performance of RF-FSO based multi-hop networks and the ones using only the RF- or the FSO-based communication. Also, the HARQ can effectively improve the energy efficiency and compensate for the effect of hardware impairments. Finally, the outage probability of multi-hop networks is not sensitive to the large number of RF-based transmit antennas while the ergodic rate is significantly affected by the number of antennas.

\vspace{-0mm}
\section*{Acknowledgement}
The research leading to these results received funding from the European Commission H2020 programme under grant agreement $n^{\circ}$671650 (5G PPP mmMAGIC project), and from the Swedish Governmental Agency for Innovation Systems (VINNOVA) within the VINN Excellence Center Chase.
\bibliographystyle{IEEEtran} 
\bibliography{masterFSO3}

\begin{thebibliography}{10}
\providecommand{\url}[1]{#1}
\csname url@samestyle\endcsname
\providecommand{\newblock}{\relax}
\providecommand{\bibinfo}[2]{#2}
\providecommand{\BIBentrySTDinterwordspacing}{\spaceskip=0pt\relax}
\providecommand{\BIBentryALTinterwordstretchfactor}{4}
\providecommand{\BIBentryALTinterwordspacing}{\spaceskip=\fontdimen2\font plus
\BIBentryALTinterwordstretchfactor\fontdimen3\font minus
  \fontdimen4\font\relax}
\providecommand{\BIBforeignlanguage}[2]{{%
\expandafter\ifx\csname l@#1\endcsname\relax
\typeout{** WARNING: IEEEtran.bst: No hyphenation pattern has been}%
\typeout{** loaded for the language `#1'. Using the pattern for}%
\typeout{** the default language instead.}%
\else
\language=\csname l@#1\endcsname
\fi
#2}}
\providecommand{\BIBdecl}{\relax}
\BIBdecl

\bibitem{6331134}
A.~Vavoulas, H.~G. Sandalidis, and D.~Varoutas, ``Weather effects on {FSO}
  network connectivity,'' \emph{IEEE J. Opt. Commun. Netw.}, vol.~4, no.~10,
  pp. 734--740, Oct. 2012.

\bibitem{6932439}
L.~Yang, X.~Gao, and M.-S. Alouini, ``Performance analysis of relay-assisted
  all-optical {FSO} networks over strong atmospheric turbulence channels with
  pointing errors,'' \emph{J. Lightw. Technol.}, vol.~32, no.~23, pp.
  4011--4018, Dec. 2014.

\bibitem{6887284}
M.~Usman, H.~C. Yang, and M.-S. Alouini, ``Practical switching-based hybrid
  {FSO/RF} transmission and its performance analysis,'' \emph{IEEE Photon. J.},
  vol.~6, no.~5, pp. 1--13, Oct. 2014.

\bibitem{4610745}
F.~Nadeem, B.~Flecker, E.~Leitgeb, M.~S. Khan, M.~S. Awan, and T.~Javornik,
  ``Comparing the fog effects on hybrid network using optical wireless and
  {GHz} links,'' in \emph{Proc. {IEEE} CNSDSP'2008}, Graz, Austria, July 2008,
  pp. 278--282.

\bibitem{1399401}
H.~Wu, B.~Hamzeh, and M.~Kavehrad, ``Achieving carrier class availability of
  {FSO} link via a complementary {RF} link,'' in \emph{Proc. {IEEE}
  Asilomar'2004}, California, USA, Nov. 2004, pp. 1483--1487.

\bibitem{4168193}
Z.~Jia, F.~Ao, and Q.~Zhu, ``{BER} performance of the hybrid {FSO/RF}
  attenuation system,'' in \emph{Proc. {IEEE} ISAPE'2006}, Guilin, China, Oct.
  2006, pp. 1--4.

\bibitem{4393998}
T.~Kamalakis, I.~Neokosmidis, A.~Tsipouras, S.~Pantazis, and I.~Andrikopoulos,
  ``Hybrid free space optical/millimeter wave outdoor links for broadband
  wireless access networks,'' in \emph{Proc. {IEEE} PIMRC'2007}, Athens,
  Greece, Sept. 2007, pp. 1--5.

\bibitem{Hamzeh}
H.~Wu, B.~Hamzeh, and M.~Kavehrad, ``Availability of airbourne hybrid {FSO/RF}
  links,'' in \emph{Proc. SPIE}, 2005, vol. 5819.

\bibitem{6364576}
Y.~Tang and M.~Brandt-Pearce, ``Link allocation, routing and scheduling of
  {FSO} augmented {RF} wireless mesh networks,'' in \emph{Proc. {IEEE}
  ICC'2012}, Ottawa, Canada, June 2012, pp. 3139--3143.

\bibitem{6400459}
A.~Sharma and R.~S. Kaler, ``Designing of high-speed inter-building
  connectivity by free space optical link with radio frequency backup,''
  \emph{IET Commun.}, vol.~6, no.~16, pp. 2568--2574, Nov. 2012.

\bibitem{6503564}
K.~Kumar and D.~K. Borah, ``Hybrid {FSO/RF} symbol mappings: Merging high speed
  {FSO} with low speed {RF} through {BICM-ID},'' in \emph{Proc. {IEEE}
  GLOBECOM'2012}, California, USA, Dec. 2012, pp. 2941--2946.

\bibitem{5342330}
N.~Letzepis, K.~D. Nguyen, A.~Guillen~i Fabregas, and W.~G. Cowley, ``Outage
  analysis of the hybrid free-space optical and radio-frequency channel,''
  \emph{IEEE J. Sel. Areas Commun.}, vol.~27, no.~9, pp. 1709--1719, Dec. 2009.

\bibitem{4411336}
S.~Vangala and H.~Pishro-Nik, ``A highly reliable {FSO/RF} communication system
  using efficient codes,'' in \emph{Proc. {IEEE} GLOBECOM'2007}, Washington,
  DC, USA, Nov. 2007, pp. 2232--2236.

\bibitem{4348339}
I.~B. Djordjevic, B.~Vasic, and M.~A. Neifeld, ``Power efficient {LDPC}-coded
  modulation for free-space optical communication over the atmospheric
  turbulence channel,'' in \emph{Proc. OFC/NFOEC'2007}, Anaheim, CA, USA, March
  2007, pp. 1--3.

\bibitem{6222288}
Y.~Tang, M.~Brandt-Pearce, and S.~G. Wilson, ``Link adaptation for throughput
  optimization of parallel channels with application to hybrid {FSO/RF}
  systems,'' \emph{IEEE Trans. Commun.}, vol.~60, no.~9, pp. 2723--2732, Sept.
  2012.

\bibitem{5351671}
B.~He and R.~Schober, ``Bit-interleaved coded modulation for hybrid {RF/FSO}
  systems,'' \emph{IEEE Trans. Commun.}, vol.~57, no.~12, pp. 3753--3763, Dec.
  2009.

\bibitem{5427418}
A.~Abdulhussein, A.~Oka, T.~T. Nguyen, and L.~Lampe, ``Rateless coding for
  hybrid free-space optical and radio-frequency communication,'' \emph{IEEE
  Trans. Wireless Commun.}, vol.~9, no.~3, pp. 907--913, March 2010.

\bibitem{6692504}
J.~Perez-Ramirez and D.~K. Borah, ``Design and analysis of bit selections in
  {HARQ} algorithm for hybrid {FSO/RF} channels,'' in \emph{Proc. IEEE VTC
  Spring'2013}, Dresden, Germany, June 2013, pp. 1--5.

\bibitem{7445896}
B.~Makki, T.~Svensson, T.~Eriksson, and M.~S. Alouini, ``On the performance of
  {RF-FSO} links with and without hybrid {ARQ},'' \emph{IEEE Trans. Wireless
  Commun.}, vol.~15, no.~7, pp. 4928--4943, July 2016.

\bibitem{6831655}
K.~Kumar and D.~K. Borah, ``Relaying in fading channels using quantize and
  encode forwarding through optical wireless links,'' in \emph{IEEE
  GLOBECOM'2013}, Atlanta, GA, USA, Dec. 2013, pp. 3741--3747.

\bibitem{6866170}
------, ``Quantize and encode relaying through {FSO} and hybrid {FSO/RF}
  links,'' \emph{IEEE Trans. Veh. Technol.}, vol.~64, no.~6, pp. 2361--2374,
  June 2015.

\bibitem{6678140}
H.~Samimi and M.~Uysal, ``End-to-end performance of mixed {RF/FSO} transmission
  systems,'' \emph{IEEE/OSA Journal of Optical Communications and Networking},
  vol.~5, no.~11, pp. 1139--1144, Nov. 2013.

\bibitem{5999707}
E.~Lee, J.~Park, D.~Han, and G.~Yoon, ``Performance analysis of the asymmetric
  dual-hop relay transmission with mixed {RF/FSO} links,'' \emph{IEEE Photon.
  Technol. Lett.}, vol.~23, no.~21, pp. 1642--1644, Nov. 2011.

\bibitem{7127443}
E.~Zedini, I.~S. Ansari, and M.~S. Alouini, ``Unified performance analysis of
  mixed line of sight {RF-FSO} fixed gain dual-hop transmission systems,'' in
  \emph{Proc. IEEE WCNC'15}, New Orleans, LA, USA, March 2015, pp. 46--51.

\bibitem{6512100}
I.~S. Ansari, F.~Yilmaz, and M.~S. Alouini, ``Impact of pointing errors on the
  performance of mixed {RF/FSO} dual-hop transmission systems,'' \emph{IEEE
  Wireless Commun. Lett.}, vol.~2, no.~3, pp. 351--354, June 2013.

\bibitem{7055847}
J.~Zhang, L.~Dai, Y.~Zhang, and Z.~Wang, ``Unified performance analysis of
  mixed radio frequency/free-space optical dual-hop transmission systems,''
  \emph{J. Lightw. Technol.}, vol.~33, no.~11, pp. 2286--2293, June 2015.

\bibitem{6775014}
N.~I. Miridakis, M.~Matthaiou, and G.~K. Karagiannidis, ``Multiuser relaying
  over mixed {RF/FSO} links,'' \emph{IEEE Trans. Commun.}, vol.~62, no.~5, pp.
  1634--1645, May 2014.

\bibitem{mmmagicdeliverable}
mmMagic Deliverable~{D1.1}, ``Use case characterization, {KPI}s and preferred
  suitable frequency ranges for future {5G} systems between 6 {GH}z and 100
  {GH}z,'' Nov. 2015, available at: https://5g-mmmagic.eu/results/.

\bibitem{ericssonAB}
{Ericsson AB}, ``Delivering high-capacity and cost-efficient backhaul for
  broadband networks today and in the future,'' Sept. 2015, available at:
  http://www.ericsson.com/res/docs/2015/microwave-2020-report.pdf.

\bibitem{mmmagic}
{European Commission 5G project}, ``{M}m-wave based mobile radio access network
  for {5G} integrated communications,'' https://5g-ppp.eu/mmmagic/.

\bibitem{mmwrician1}
H.~Xu, T.~S. Rappaport, R.~J. Boyle, and J.~H. Schaffner, ``Measurements and
  models for {38-GHz} point-to-multipoint radiowave propagation,'' \emph{IEEE
  J. Sel. Areas Commun.}, vol.~18, no.~3, pp. 310--321, March 2000.

\bibitem{mmwrician2}
D.~Beauvarlet and K.~L. Virga, ``Measured characteristics of {30-GHz} indoor
  propagation channels with low-profile directional antennas,'' \emph{IEEE
  Antennas Wireless Propag. Lett.}, vol.~1, no.~1, pp. 87--90, 2002.

\bibitem{mmwrician3}
A.~Annamalai, G.~Deora, and C.~Tellambura, ``Analysis of generalized selection
  diversity systems in wireless channels,'' \emph{IEEE Trans. Veh. Technol.},
  vol.~55, no.~6, pp. 1765--1775, Nov. 2006.

\bibitem{phdthesisBjornemo}
E.~Bjornemo, ``Energy constrained wireless sensor networks: communication
  principles and sensing aspects,'' {P}h.D. dissertation, Uppsala University,
  Uppsala, Sweden, 2009.

\bibitem{6515206}
D.~Persson, T.~Eriksson, and E.~G. Larsson, ``Amplifier-aware multiple-input
  multiple-output power allocation,'' \emph{IEEE Commun. Lett.}, vol.~17,
  no.~6, pp. 1112--1115, June 2013.

\bibitem{6725577}
------, ``Amplifier-aware multiple-input single-output capacity,'' \emph{IEEE
  Trans. Commun.}, vol.~62, no.~3, pp. 913--919, March 2014.

\bibitem{7104158}
B.~Makki, T.~Svensson, T.~Eriksson, and M.~Nasiri-Kenari, ``On the throughput
  and outage probability of multi-relay networks with imperfect power
  amplifiers,'' \emph{IEEE Trans. Wireless Commun.}, vol.~14, no.~9, pp.
  4994--5008, Sept. 2015.

\bibitem{throughputdef}
G.~Caire and D.~Tuninetti, ``The throughput of hybrid-{ARQ} protocols for the
  {G}aussian collision channel,'' \emph{IEEE Trans. Inf. Theory}, vol.~47,
  no.~5, pp. 1971--1988, July 2001.

\bibitem{MIMOARQkhodemun}
B.~Makki and T.~Eriksson, ``On the performance of {MIMO-ARQ} systems with
  channel state information at the receiver,'' \emph{IEEE Trans. Commun.},
  vol.~62, no.~5, pp. 1588--1603, May 2014.

\bibitem{a01661837}
H.~E. Gamal, G.~Caire, and M.~O. Damen, ``The {MIMO ARQ} channel:
  Diversity-multiplexing-delay tradeoff,'' \emph{IEEE Trans. Inf. Theory},
  vol.~52, no.~8, pp. 3601--3621, Aug. 2006.

\bibitem{tuninetti2011}
D.~Tuninetti, ``On the benefits of partial channel state information for
  repetition protocols in block fading channels,'' \emph{IEEE Trans. Inf.
  Theory}, vol.~57, no.~8, pp. 5036--5053, Aug. 2011.

\bibitem{6168189}
S.~M. Aghajanzadeh and M.~Uysal, ``Information theoretic analysis of
  hybrid-{ARQ} protocols in coherent free-space optical systems,'' \emph{IEEE
  Trans. Commun.}, vol.~60, no.~5, pp. 1432--1442, May 2012.

\bibitem{wolframwebsite}
{Wolfram}, ``The wolfram functions site,'' [Online], Available:
  http://functions.wolfram.com, 2016.

\bibitem{bookhypergeometric}
I.~S. Gradshteyn and I.~M. Ryzhik, \emph{Table of Integrals, Series, and
  Products}.\hskip 1em plus 0.5em minus 0.4em\relax 7th edition. Academic
  Press, San Diego, 2007.

\bibitem{5357980}
F.~Yilmaz and M.-S. Alouini, ``Product of shifted exponential variates and
  outage capacity of multicarrier systems,'' in \emph{Proc. European Wireless},
  Alborg, Denmark, May 2009, pp. 282--286.

\bibitem{minkowskibook}
R.~A. Horn and C.~R. Johnson, \emph{Matrix Analysis}.\hskip 1em plus 0.5em
  minus 0.4em\relax Cambridge University Press, 1985.

\end{thebibliography}
\vfill
\end{document}